%% file: Article-arXiv.tex
\documentclass[10pt, prd, reprint, onecolumn, nofootinbib, showpacs]{revtex4-1}

\usepackage[utf8]{inputenc}

\usepackage{amsmath}

\usepackage{amsfonts}
\usepackage{mathrsfs}

\usepackage{amssymb}
\usepackage{units}

\usepackage{graphicx}

\usepackage{placeins}

\usepackage{verbatim}

\usepackage{accents}

\usepackage{multirow}

\usepackage{color}

\newcommand{\mrm}{\mathrm}

\makeatletter

\let\old@dmathbeg\[
\let\old@dmathend\]

\newcommand{\rovnec}[1]{\old@dmathbeg#1\old@dmathend}
\newcommand{\rovcis}[2]{\begin{equation}#1\label{#2}\end{equation}}
\newcommand{\drovcis}[2]{\begin{equation}\begin{split}#1\end{split}\label{#2}\end{equation}}
\newcommand{\drovnec}[1]{\begin{equation*}\begin{split}#1\end{split}\end{equation*}} 
\newcommand{\provcis}[1]{\begin{align}#1\end{align}}
\newcommand{\provnec}[1]{\begin{align*}#1\end{align*}}

\newcommand{\rov}{\@ifstar\rovnec\rovcis}
\newcommand{\drov}{\@ifstar\drovnec\drovcis}
\newcommand{\prov}{\@ifstar\provnec\provcis}

\newcommand{\vast}{\bBigg@{4}}
\newcommand{\Vast}{\bBigg@{5}}

\newcommand{\setstretch}[1]{%
  \def\baselinestretch{#1}%
  \@currsize
}

\makeatother

\DeclareMathOperator{\sgn}{sgn}

\DeclareMathOperator{\diffbold}{\mathbf{d}}
\newcommand{\bd}{\diffbold\!}

\newcommand{\msc}{\mathscr}
\newcommand{\mbs}{\boldsymbol}

\newcommand{\iDelta}{{\mit\Delta}}

\newcommand{\iSigma}{{\mit\Sigma}}

\newcommand{\iPi}{{\mit\Pi}}

\DeclareFontEncoding{LGR}{}{}

\DeclareMathAlphabet{\mgr}{LGR}{cmr}{m}{n}

\newcommand{\rpi}{\mgr{p}}

\newcommand{\equival}{\Longleftrightarrow}

\DeclareMathOperator{\conj}{\&}

\renewcommand{\[}{\left[}
\renewcommand{\]}{\right]}

\newcommand{\f}{\!\left}
\newcommand{\ri}{\right} 

\newcommand{\der}[3][]{\frac{\mrm{d}^{#1}#2}{\mrm{d}{#3}^{#1}}}
\newcommand{\pder}[3][]{\frac{\partial^{#1}#2}{\partial{#3}^{#1}}}
\newcommand{\dpder}[2]{\pder[2]{#1}{#2}}

\newcommand{\bpd}[2][]{\frac{\boldsymbol\partial #1}{\boldsymbol\partial #2}} 
\newcommand{\nicepder}[3][]{\nicefrac{\partial^{#1}#2}{\partial{#3}^{#1}}}
\newcommand{\nicedpder}[2]{\nicepder[2]{#1}{#2}}
\newcommand{\nicebpd}[2][]{\nicefrac{\boldsymbol\partial #1}{\boldsymbol\partial #2}}

\newcommand{\zrov}{{}\\{}}

\newcommand{\res}[2]{\left.#1\ri|_{#2}}
\newcommand{\lbl}{\label}
\newcommand{\rvt}{\ .}
\newcommand{\rvc}{\ ,}
\newcommand{\rvs}{\ ;}
\newcommand{\qt}[1]{``#1''}

\renewcommand{\(}{\left(}
\renewcommand{\)}{\right)}

\begin{document}

\title{Kinematic restrictions on particle collisions near extremal black holes:\\ A unified picture}

\author{Filip Hejda}
\email{filip.hejda@tecnico.ulisboa.pt}
\affiliation{Centro Multidisciplinar de Astrofísica -- CENTRA, Departamento de
Física, Instituto Superior Técnico -- IST, Universidade de Lisboa -- UL, Avenida Rovisco Pais 1, 1049-001 Lisboa, Portugal}
\affiliation{Institute of Theoretical Physics, Faculty of Mathematics and Physics,
Charles University,
V Holešovičkách 2, 180\,00 Prague 8, Czech Republic}
\author{Jiří Bičák}
\email{bicak@mbox.troja.mff.cuni.cz}
\affiliation{Institute of Theoretical Physics, Faculty of Mathematics and Physics,
Charles University,
V Holešovičkách 2, 180\,00 Prague 8, Czech Republic}

\pacs{04.20.--q, 04.20.Jb, 04.40.Nr, 04.70.Bw}

\begin{abstract}
In 2009, Bañados, Silk and West (BSW) pointed out the possibility of having an unbounded limit of centre-of-mass collision energy for test particles in the field of an extremal Kerr black hole, if one of them has fine-tuned parameters and the collision point is approaching the horizon. The possibility of this \qt{BSW effect} attracted much attention: it was generalised to arbitrary (\qt{dirty}) rotating black holes and an analogy was found for collisions of charged particles in the field of non-rotating charged black holes. Our work considers the unification of these two mechanisms, which have so far been studied only separately. Exploring the enlarged parameter space, we find kinematic restrictions that may prevent the fine-tuned particles from reaching the limiting collision point. These restrictions are first presented in a general form, which can be used with an arbitrary black-hole model, and then visualised for the Kerr-Newman solution by plotting the \qt{admissible region} in the parameter space of critical particles, reproducing some known results and obtaining a number of new ones. For example, we find that (marginally) bounded critical particles with enormous values of angular momentum can, curiously enough, approach the degenerate horizon, if the charge of the black hole is very small. Such \qt{mega-BSW} behaviour is excluded in the case of a vacuum black hole, or a black hole with large charge. It may be interesting in connection with the small \qt{Wald charge} induced on rotating black holes in external magnetic fields.
\end{abstract}

\maketitle

\tableofcontents

\section{Introduction}

After initial discussions of how to extract energy from a black hole by particle decays in its vicinity \cite{Penrose69,Christd70}, a concern was raised whether this \qt{Penrose process} is compatible with realistic interactions responsible for the disruption \cite{BarPrTeu}. Further studies \cite{PirShK, PirSh} sought more possibilities, turning to much less restrictive collision kinematics (the \qt{collisional Penrose process}). One of their observations was that in certain limiting cases the centre-of-mass collision energy between the colliding particles can grow unboundedly. This gained considerable attention much later, after Bañados, Silk and West (BSW) \cite{BSW} discovered another, more subtle yet more realistic, possibility of this kind. In this so-called BSW effect a collision between two ingoing particles in the field of the extremal Kerr black hole is considered. One of the particles must have fine-tuned parameters (angular momentum), whereas the other must not. Then, the centre-of-mass collision energy can grow without bound if the collision point is approaching the black-hole horizon.

Various other scenarios of obtaining arbitrarily high centre-of-mass collision energy were explored in subsequent works. For instance, in the Piran-Shaham effect (called so retrospectively after \cite{PirSh}), a collision of an outgoing and ingoing particle near an extremal black hole is considered, see e.g. \cite{Zasl11b}. For a general review, cf. \cite{HK14}. Some other examples are mentioned in the discussion of collisions near the inner horizon in \cite{Zasl12a}. Multiple-scattering mechanisms are also frequently studied. In scenarios where the particles taking part in the final collision cannot reach the spot from far away due to kinematic restrictions, they can be produced in a preceding collision. This idea was pioneered by Grib and Pavlov \cite{GP10a, GP11a} using an astrophysically relevant setup (a fast-spinning but non-extremal Kerr black hole).

The proposal by Grib and Pavlov can be seen partly as a response to ongoing disputes about the practical limitations of the BSW effect. In particular, the maximum achievable centre-of-mass energy with particles coming from rest at infinity grows very slowly with the decreasing deviation from extremality \cite{BCGPU, JacSot}. Further issue with the BSW-like processes is the significance of various types of back-reaction. For example, gravitational wave emission related to the BSW effect was modeled in \cite{BCGPU}. A different aspect of the problem is the adequacy of the test particle approximation for these processes. Kimura \emph{et al.} \cite{KimNakTag} turned to colliding dust shells in Kerr background, so that they could include self-gravity of the colliding objects. In their setup, the collision energy observable from infinity turned out to be finite. On the other hand, later results indicate that the BSW effect still exists for point particles even when some self-force contributions are considered \cite{TantZasl}. Another issue is whether the high-energy collisions can lead to a significant energy extraction and observable signatures. The stringent early assessment of bounds on extracted energy in \cite{JacSot} was identified as too crude in \cite{GP10a}. Later, more detailed investigations of energy extraction were conducted \cite{HaNeMi, Zasl12b, Zasl12c, BPAH, Schnitt14, BeBrCa, Zasl15b}, sometimes using the term \qt{super-Penrose process} for the possible high-energy gain. Observable signatures of high-energy collisions (annihilations) of conjectured dark-matter particles were estimated in \cite{BHSW, Schnitt15}. 

Regardless of all complexities, even the original BSW-type effect within the test-particle regime is an interesting theoretical issue and its investigation has continued. Harada and Kimura analysed the BSW effect for general, non-equatorial particles in the Kerr field \cite{HK11b}, finding that it does not work around poles. These results were extended in \cite{Zasl12d, Liu13}. Other enhancements stayed within equatorial motion, two of which we find especially important. First, the BSW-type effect was generalised to arbitrary rotating black holes by Zaslavskii \cite{Zasl10}. He also noted that fine-tuning of the parameters of one of the particles is not sufficient for the BSW effect to occur, since there can be further restrictions, which are (unlike the fine-tuning) model dependent. A particular example is given in \cite{WeiLiuGuoFu}, where it is shown that there exists a minimum value of the spin parameter of the extremal Kerr-Newman solution for the BSW effect (with uncharged particles) to be possible. Second, Zaslavskii also found \cite{Zasl11a} that there is an analogy of the BSW-type effect with charged particles in the Reissner-Nordström spacetime, which occurs even for radial motion. 
 
These two variants of the generalised BSW effect have been studied separately. More importantly, there does not appear to be any study systematically discussing the additional restrictions and, for example, generalising the results of \cite{WeiLiuGuoFu} for the Kerr-Newman solution to charged particles. We deliver such a study in the present paper: our purpose is to examine kinematic restrictions for the generalised BSW effect including effects of \emph{both} dragging and electrostatic interaction. We base our study on the general metric form, which can include black holes with different types of (matter) fields, sometimes called \qt{dirty} black holes, or black holes in spacetimes which are not asymptotically flat.

The generalised BSW effect always constitutes a \qt{corner case} of the test-particle kinematics, and considering the setup with both charge and dragging requires a further increased rigour. Moreover, the notation and methods vary significantly among different authors. Thus, to be able to give a unified picture, in Section \ref{sek:egm} we thoroughly go through methods of qualitative study of electrogeodesic motion, building up on classical works of Wilkins \cite{Wilkins} and Bardeen \cite{Bardeen72}. Some further details are given in Appendix \ref{app:derwveff}.

In Section \ref{sek:col} we review how to take a horizon limit of the centre-of-mass collision energy and the way to show that there are particles with distinct type of motion in a near-horizon zone, so-called critical particles, and that these particles cause the BSW singularity in the centre-of-mass collision energy. We present formulae for different types of collisions. 

Section \ref{sek:kincr} contains the main results. We prove that the critical particles can approach the position of the horizon only if it is degenerate and their parameters satisfy certain restrictions. We discuss how these restrictions depend on the properties of the black hole and identify two cases corresponding to the original centrifugal and electrostatic mechanisms of the generalised BSW effect. Two other \qt{mixed} cases are also seen to be possible.

In Section \ref{sek:kn} we illustrate these results with the example of the extremal Kerr-Newman solution, where just one of the mixed cases applies. Apart from the general restrictions on particles at \emph{any} energy, we study what happens for the critical particles that are coming from rest at infinity or are bounded. We notice that, for a very small charge of the black hole, this kind of particle can reach the position of the degenerate horizon even with enormous values of angular momentum (and specific charge). Such a \qt{mega-BSW} effect is possible neither in the vacuum case nor in the case with a large charge of the black hole.

Finally, let us mention one area where the present work can be extended. As we discussed in the introduction of \cite{Article1}, the interaction of black holes with external (magnetic) fields is of considerable astrophysical interest, even for (nearly) extremal black holes. Particle collisions near magnetised black holes have been already studied, first in a simple model by Frolov \cite{Frolov12} and later by others in more versatile setups \cite{IgHaKi, Zasl15a}. However, these works considered only test fields. In contrast, in \cite{Article1} we studied exact models of black holes in strong external magnetic fields. Combining the presented general scheme for examining particle collisions with results and techniques of \cite{Article1}, we hope to get a new perspective on the problem.\footnote{The paper is based on preliminary results obtained in \cite{dipl}, which were substantially reworked and augmented during the Ph.D. study of F.H. at CENTRA in Lisbon.}

\section{Electrogeodesic motion in black-hole spacetimes}

\lbl{sek:egm}

Let us consider an axially symmetric, stationary spacetime with the metric
\rov{\mbs g=-N^2\bd t^2+g_{\varphi\varphi}\(\bd\varphi-\omega\bd t\)^2+g_{rr}\bd r^2+g_{\vartheta\vartheta}\bd\vartheta^2\rvc}{axst}
which will serve as a model of an electrovacuum black hole. (The cosmological constant can also be included. For conditions on matter fields compatible with \eqref{axst}, see e.g. \cite{FrolNov} and references therein.) 

We assume the choice of coordinate $r$ such that hypersurface $r=r_+$ is the black-hole horizon, where $N^2$ vanishes. If the black-hole horizon is degenerate (which will be denoted by $r_0$), the following factorisation of \eqref{axst} is useful: 
\rov{\mbs g=-\(r-r_0\)^2{\tilde N}^2\bd t^2+g_{\varphi\varphi}\(\bd\varphi-\omega\bd t\)^2+\frac{\tilde g_{rr}}{(r-r_0)^2}\bd r^2+g_{\vartheta\vartheta}\bd\vartheta^2\rvc}{axstex}
where $\tilde N^2$ and $\tilde g_{rr}$ are non-vanishing and finite at $r=r_0$.

Let us further require that the electromagnetic field accompanying \eqref{axst} has the same symmetry as the metric, exhibited by the following choice of gauge:
\rov{\mbs A=A_t\bd t+A_\varphi\bd\varphi=-\phi\bd t+A_\varphi\(\bd\varphi-\omega\bd t\)\rvt}{aaxst}
Here $\phi=-A_t-\omega A_\varphi$ will be called the generalised electrostatic potential. Introducing the locally non-rotating frame (cf. \cite{BarPrTeu}) associated with \eqref{axst},
\prov{\mbs e_{(t)}&=\frac{1}{N}\(\bpd{t}+\omega\bpd{\varphi}\)\rvc&\mbs e_{(\varphi)}&=\frac{1}{\sqrt{g_{\varphi\varphi}}}\bpd{\varphi}\rvc\lbl{lnrf1}\\\mbs e_{(r)}&=\frac{1}{\sqrt{g_{rr}}}\bpd{r}\rvc&\mbs e_{(\vartheta)}&=\frac{1}{\sqrt{g_{\vartheta\vartheta}}}\bpd{\vartheta}\rvc\lbl{lnrf2}}
we see that $\phi$ is proportional to $A_{(t)}$. Let us also consider an energy of a test particle locally measured in this frame given by $\varepsilon_\mrm{LNRF}\equiv u^{(t)}$.

Equations of motion for test particles (with rest mass $m$ and charge $q=\tilde qm$) influenced solely by the Lorentz force, i.e. of electrogeodesic motion,  can be derived from the Lagrangian 
\rov{\msc L=\frac{1}{2}mg_{\mu\nu}u^\mu u^\nu+qA_\mu u^\mu\rvt}{eglag}
The corresponding canonical momentum is
\rov{\iPi_\alpha=\pder{\msc L}{u^\alpha}=p_\alpha+qA_\alpha\rvt}{}
Its projections on $\nicebpd{t},\nicebpd{\varphi}$, the two Killing vectors of \eqref{axst}, are conserved during the electrogeodesic motion. They can be interpreted as (minus) energy $E$ and axial angular momentum $L_z$. In our coordinates they read
\prov{-\iPi_t=-p_t-qA_t&=E=\varepsilon m\rvc&\iPi_\varphi=p_\varphi+qA_\varphi&=L_z=lm\rvt}
Dividing by the mass $m$ of the particle, we can get expressions for two contravariant components of the velocity,
\prov{u^t&=\frac{\varepsilon-\omega l-\tilde q\phi}{N^2}\rvc&u^\varphi&=\frac{\omega}{N^2}\(\varepsilon-\omega l-\tilde q\phi\)+\frac{l-\tilde qA_\varphi}{g_{\varphi\varphi}}\rvt\lbl{utphgen}}
Assuming further that metric \eqref{axst} is invariant under reflections $\vartheta\to\rpi-\vartheta$, i.e. under \qt{mirror symmetry} with respect to the equatorial \qt{plane}, we can consider motion confined to this hypersurface (with conserved conditions $\vartheta=\nicefrac{\rpi}{2},u^\vartheta=0$). The remaining component of the velocity then follows from its normalisation,
\rov{u^r=\pm\sqrt{\frac{1}{N^2g_{rr}}\[\(\varepsilon-\omega l-\tilde q\phi\)^2-N^2\(1+\frac{\(l-\tilde qA_\varphi\)^2}{g_{\varphi\varphi}}\)\]}\rvt}{urgen}
Hence, we have a full set of the first-order equations of motion for an equatorial electrogeodesic test particle. (See \cite{Sharp} for references.)

There are some qualitative features of the motion that follow directly from the equations above. The motion of particles with some parameters may be forbidden in certain ranges of $r$. The first restriction comes from the conventional requirement (positivity of the locally measured energy $\varepsilon_\mrm{LNRF}$) for motion \qt{forward} in coordinate time $t$, which applies for $r>r_+$ (or $N^2>0$). From $u^t$ in \eqref{utphgen} we infer 
\rov{\msc X\equiv\varepsilon-\omega l-\tilde q\phi>0\rvt}{fwpos}
(Later, we also consider the possibility $\msc X\to0$ for $N\to0$.) 

Another restriction is due to the square root in \eqref{urgen}. If we assume that the metric determinant for \eqref{axst}, which is given by \rov{\sqrt{-g}=N\sqrt{g_{rr}g_{\varphi\varphi}g_{\vartheta\vartheta}}\rvc}{}
is non-degenerate, the expression $N^2g_{rr}$ under the square root in \eqref{urgen} is non-vanishing and positive. Therefore, the square root will be defined in real numbers if
\rov{W\equiv\(\varepsilon-\omega l-\tilde q\phi\)^2-N^2\(1+\frac{\(l-\tilde qA_\varphi\)^2}{g_{\varphi\varphi}}\)\geqslant0\rvt}{weffpos}
Zeroes of function $W$ with respect to radius are turning points (because of that, $W$ is often called effective potential). Stationary and inflection points of $W$ with respect to $r$ are used to find circular orbits and marginally stable circular orbits \cite{Wald, BarPrTeu} (see also \ref{app:derwveff}). However, $W$ is not unique; for example, if we multiply it with a positive integer power of $r$, the results will be the same. This led to different conventions in literature \cite{Wald, BarPrTeu}. Nevertheless, we can define another effective potential which will be unique and also have other advantages. 

First, for $r\geqslant r_+$, we can factorise $W$ as
\rov{W=\(\varepsilon-V_+\)\(\varepsilon-V_-\)\rvc}{wveff}
where\footnote{Note that we assume $g_{\varphi\varphi}>0$, i.e. the absence of closed timelike curves.}
\rov{V_\pm=\omega l+\tilde q\phi\pm N\sqrt{1+\frac{\(l-\tilde qA_\varphi\)^2}{g_{\varphi\varphi}}}\rvt}{veffgen}
Since $V_+\geqslant V_-$, the condition $W\geqslant 0$ is fulfilled whenever $\varepsilon\geqslant V_+$ or $\varepsilon\leqslant V_-$. Considering $l=0,\tilde q=0$ and comparing with \eqref{fwpos}, we can identify $\varepsilon\leqslant V_-$ as the domain of unphysical particles moving \qt{backwards in time}. (Conversely, we see that restriction $\varepsilon\geqslant V_+$ is stronger than \eqref{fwpos}, so it ensures motion \qt{forward in time}, and manifests \eqref{fwpos} to be preserved during motion.) 

Therefore, we can define $V\equiv V_+$ and use $\varepsilon\geqslant V$ as a condition for the motion to be allowed (in the $r\geqslant r_+$, or $N^2\geqslant0$, domain). In this sense, $V$ is the best analogy of a classical effective potential. It is also called a \qt{minimum energy} \cite{Bardeen72}. The ranges of radii where $\varepsilon<V$ are referred to as \qt{forbidden bands}, whereas the ones with $\varepsilon>V$ are called \qt{allowed bands}. Condition $\varepsilon=V$ implies $W=0$ and thus corresponds to a turning point. For the convenience of the reader, in \ref{app:derwveff} we derive relations between the derivatives of $W$ and $V$.

\section{Collision energy and critical particles}

\lbl{sek:col}

Let us consider two colliding (charged) particles in an arbitrary spacetime. The natural generalisation of the centre-of-mass frame from special relativity is a tetrad, where the total momentum of the colliding particles at the instant of collision has just the time component
\rov{\(E_\mrm{CM},0,0,0\)=m_1\mbs u_{(1)}+m_2\mbs u_{(2)}\rvt}{}
This tetrad component can be interpreted as the centre-of-mass collision energy. To get rid of the frame, we can take square of the above expression and define an invariant related to this quantity (cf. \cite{GP11a}, for example)
\rov{\frac{E^2_\mrm{CM}}{2m_1m_2}=\frac{m_1}{2m_2}+\frac{m_2}{2m_1}-g_{\iota\kappa}u_{(1)}^\iota u_{(2)}^\kappa\rvt}{}
Let us now investigate how this invariant behaves for collisions of electrogeodesic particles in black-hole spacetimes. Using the metric coefficients of \eqref{axst}, the expressions for components of particles' velocities given by first-order equations of equatorial electrogeodesic motion \eqref{utphgen} and \eqref{urgen}, and the definition \eqref{fwpos} of \qt{forwardness} $\msc X$, we obtain
\drov{\frac{E_\mrm{CM}^2}{2m_1m_2}={}&\frac{m_1}{2m_2}+\frac{m_2}{2m_1}-\frac{\(l_1-\tilde q_1A_\varphi\)\(l_2-\tilde q_2A_\varphi\)}{g_{\varphi\varphi}}+\frac{\msc X_1\msc X_2}{N^2}\mp\zrov&\mp\frac{1}{N^2}\sqrt{\msc X_1^2-N^2\[1+\frac{\(l_1-\tilde q_1A_\varphi\)^2}{g_{\varphi\varphi}}\]}\sqrt{\msc X_2^2-N^2\[1+\frac{\(l_2-\tilde q_2A_\varphi\)^2}{g_{\varphi\varphi}}\]}\rvt}{ecmgen}
The $\mp$ sign before the last term corresponds to particles moving in the same or the opposite direction in $r$. 

Now, let us consider the limit $N\to0$. We need to Taylor expand the square roots $\sqrt{W}$ coming from the radial components of the particles' velocities. For each of the colliding particles, there are two very different cases depending on the value $\msc{X}_\mrm{H}$ of $\msc X$ on the horizon. For a generic particle ($\msc{X}_\mrm{H}>0$), the expansion looks as follows:
\rov{\sqrt{W}\doteq \msc X-\frac{N^2}{2\msc X}\[1+\frac{\(l-\tilde qA_\varphi\)^2}{g_{\varphi\varphi}}\]+\dots}{sqwus}
If we consider two generic particles moving in the same direction (upper sign in \eqref{ecmgen}), this behaviour leads to the cancellation of the terms that are singular in the limit $N\to0$, and a finite limit arises,\footnote{The case of the particles going in opposite directions, i.e. plus sign in \eqref{ecmgen}, leads to the so-called Piran-Shaham effect (cf. the Introduction).}
\drov{\res{\frac{E_\mrm{CM}^2}{2m_1m_2}}{N=0}={}&\frac{m_1}{2m_2}+\frac{m_2}{2m_1}-\res{\frac{\(l_1-\tilde q_1A_\varphi\)\(l_2-\tilde q_2A_\varphi\)}{g_{\varphi\varphi}}}{N=0}+\zrov&+\frac{1}{2}\res{\[1+\frac{\(l_2-\tilde q_2A_\varphi\)^2}{g_{\varphi\varphi}}\]}{N=0}\frac{\msc X_1^\mrm{H}}{\msc X_2^\mrm{H}}+\frac{1}{2}\res{\[1+\frac{\(l_1-\tilde q_1A_\varphi\)^2}{g_{\varphi\varphi}}\]}{N=0}\frac{\msc X_2^\mrm{H}}{\msc X_1^\mrm{H}}\rvt}{ecmgenhor}
The presence of $\msc X_\mrm{H}$ for both particles in the denominators suggests that for the so-called \qt{critical particles} with $\msc X_\mrm{H}=0$ (so far excluded, see \eqref{fwpos}) the limit may not be finite. 

To verify this, let us first expand \qt{forwardness} $\msc X$ of a critical particle around $r_+$,
\rov{\msc X\doteq-\res{\(\pder{\omega}{r}l+\tilde q\pder{\phi}{r}\)}{r=r_+,\vartheta=\frac{\rpi}{2}}\(r-r_+\)+\dots}{fwcrit}
Thus, for a critical particle, $\msc X^2$ is proportional to $\(r-r_+\)^2$ (with higher-order corrections). However, for a subextremal black hole, we expect $N^2$ to be proportional just to $r-r_+$, so the positive term under the square root in $\sqrt{W}$ will go to zero faster than the negative one. We thus anticipate that the motion of critical particles towards the horizon is forbidden for subextremal black holes. We return to these kinematic restrictions below. 

In case of an extremal black hole \eqref{axstex} with $N^2=\(r-r_0\)^2\tilde N^2$, we get an expansion for $\sqrt{W}$ of a critical particle very different from \eqref{sqwus},
\rov{\res{\sqrt{W}}{\msc X_\mrm{H}=0}\doteq\(r-r_0\)\res{\sqrt{\(\pder{\omega}{r}l+\tilde q\pder{\phi}{r}\)^2-{\tilde N}^2\[1+\frac{\(l-\tilde qA_\varphi\)^2}{g_{\varphi\varphi}}\]}}{r=r_0,\vartheta=\frac{\rpi}{2}}+\dots}{sqwcr}
Now, if we again consider particles moving in the same direction, but assume that particle $1$ is critical, whereas particle $2$ is generic (usually referred to as \qt{usual} in literature), we get the following leading-order behaviour in the limit $r\to r_0$ (or $N\to0$)
\rov{\frac{E_\mrm{CM}^2}{2m_1m_2}\approx-\frac{\msc X_2^\mrm{H}}{r-r_0}\left\{\frac{1}{{\tilde N}^2}\[\pder{\omega}{r}l_1+\tilde q_1\pder{\phi}{r}+\res{\sqrt{\(\pder{\omega}{r}l_1+\tilde q_1\pder{\phi}{r}\)^2-{\tilde N}^2\[1+\frac{\(l_1-\tilde q_1A_\varphi\)^2}{g_{\varphi\varphi}}\]}\]\ri\}}{r=r_0,\vartheta=\frac{\rpi}{2}}\rvt}{}
The expression diverges like $\(r-r_0\)^{-1}$, so we confirmed that the different behaviour of $\sqrt{W}$ for critical particles leads to singularity in the centre-of-mass collision energy invariant. 

Since the divergent contribution is proportional to $\msc X_\mrm{H}$ of the usual particle, we see that for a collision of two critical particles the limit is again finite, namely
\drov{\res{\frac{E_\mrm{CM}^2}{2m_1m_2}}{N=0}={}&\frac{m_1}{2m_2}+\frac{m_2}{2m_1}-\res{\frac{\(l_1-\tilde q_1A_\varphi\)\(l_2-\tilde q_2A_\varphi\)}{g_{\varphi\varphi}}}{r=r_0,\vartheta=\frac{\rpi}{2}}+\zrov&+\res{\left\{\frac{1}{{\tilde N}^2}\[\(\pder{\omega}{r}l_1+\tilde q_1\pder{\phi}{r}\)\(\pder{\omega}{r}l_2+\tilde q_2\pder{\phi}{r}\)\mp\sqrt{\frac{1}{2}\pder[2]{W_{(1)}}{r}}\sqrt{\frac{1}{2}\pder[2]{W_{(2)}}{r}}\]\ri\}}{r=r_0,\vartheta=\frac{\rpi}{2}}\rvc}{}
where we observed 
\rov{\res{\pder[2]{W}{r}}{r=r_0,\msc X_\mrm{H}=0}=2\res{\left\{\(\pder{\omega}{r}l+\tilde q\pder{\phi}{r}\)^2-{\tilde N}^2\[1+\frac{\(l-\tilde qA_\varphi\)^2}{g_{\varphi\varphi}}\]\ri\}}{r=r_0,\vartheta=\frac{\rpi}{2}}\rvt}{}

\section{Kinematics of critical particles}

\lbl{sek:kincr}

We have seen that particles with zero forwardness $\msc X$ at the horizon, i.e. critical particles, constitute a distinct kind of motion with a different behaviour of the radial velocity, leading to the singularity in the collision energy. The condition $\msc X_\mrm{H}=0$ can be formulated as a requirement for the particle's energy to have a critical value 
\rov{\varepsilon_\mrm{cr}=l\omega_\mrm{H}+\tilde q\phi_\mrm{H}=\res{V}{r_+}\rvc}{critgen}
which coincides with the value of effective potential \eqref{veffgen} at the radius of the horizon. Thus if the minimum energy $V$ grows for $r>r_+$, the motion of critical particles towards $r_+$ is forbidden, since their energy will be lower than that allowed for $r>r_+$. On the other hand, if the effective potential decreases, the motion of critical particles towards $r_+$ will be allowed. Thus, we have to look at the sign of the radial derivative of $V$ to discriminate between the cases. For the geodesic ($\tilde q=0$) case, the discussion has  already been carried out by Zaslavskii \cite{Zasl10}, who utilised rather mathematical considerations contained in \cite{Medved}.

\subsection{Derivative of the effective potential}

Taking the derivative of the effective potential \eqref{veffgen}, we obtain four terms
\rov{\pder{V}{r}=\pder{\omega}{r}l+\tilde q\pder{\phi}{r}+\pder{N}{r}\sqrt{1+\frac{\(l-\tilde qA_\varphi\)^2}{g_{\varphi\varphi}}}+N\pder{}{r}\(\sqrt{1+\frac{\(l-\tilde qA_\varphi\)^2}{g_{\varphi\varphi}}}\)\rvt}{}
The fourth term is proportional to $N$, so we do not consider it in the limit $N\to0$. The third term can be modified to the following form:
\rov{\lim_{N\to0}\pder{V}{r}=\lim_{N\to0}\res{\[\pder{\omega}{r}l+\tilde q\pder{\phi}{r}+\frac{\pder{\(N^2\)}{r}}{2N}\sqrt{1+\frac{\(l-\tilde qA_\varphi\)^2}{g_{\varphi\varphi}}}\]}{\vartheta=\frac{\rpi}{2}}\rvt}{}
In subextremal cases the radial derivative of $N^2$ is nonzero for $N\to0$, i.e. the third term blows up in the limit that we wish to take. This term is manifestly positive in the near-horizon regime, so its domination means that no critical particles can approach $r_+$ for subextremal black holes. Zaslavskii's result \cite{Zasl10} is thus generalised to the $\tilde q\neq0$ case.

On the other hand, in the extremal case the radial derivative of $N^2$ vanishes in the limit $N\to0$. Using again the decomposition \eqref{axstex}, $N^2=\(r-r_0\)^2\tilde N^2$, we get (for $r\geqslant r_0$) $\nicepder{N}{r}=\tilde N+\(r-r_0\)\nicepder{\tilde N}{r}$. 
This enables us to take the $r\to r_0$ limit of the third term and to drop the contribution proportional to $r-r_0$; thus,
\rov{\res{\pder{V}{r}}{r=r_0}=\res{\(\pder{\omega}{r}l+\tilde q\pder{\phi}{r}+\tilde N\sqrt{1+\frac{\(l-\tilde qA_\varphi\)^2}{g_{\varphi\varphi}}}\)}{r=r_0, \vartheta=\frac{\rpi}{2}}\rvt}{veffderh}
This final expression is indeed finite. However, it depends heavily on the parameters $l,\tilde q$ of the particle as well on the properties of the black-hole model in question. Thus (as already noted by Zaslavskii in \cite{Zasl10} for the $\tilde q=0$ case) kinematic restrictions on the motion of critical particles towards $r_0$ cannot be worked out in a model-independent way. However, one can qualitatively study the dependence of the kinematic restrictions on the features of a general model and then use these considerations to get quantitative results for particular models. This is our main aim in what follows.

\subsection{Remarks on motion towards $r_0$}

\lbl{odd:motcr}

Before we analyse when the motion of critical particles towards $r_0$ is allowed and when it is not, let us first elucidate some features that this motion has if it is allowed. First, let us note that, comparing the behaviour of $\msc X$ \eqref{fwcrit} and $N^2=\(r-r_0\)^2\tilde N^2$, we see that $u^t\to\infty$ with $r\to r_0$ even for critical particles. For usual particles the locally measured energy $\varepsilon_\mrm{LNRF}$ also blows up. However, the $r\to r_0, r\geqslant r_0$ limit of $\varepsilon_\mrm{LNRF}$ for the critical particles is finite, namely,
\rov{\varepsilon_\mrm{LNRF}\equiv u^{(t)}\doteq-\res{\frac{\pder{\omega}{r}l+\tilde q\pder{\phi}{r}}{\tilde N}}{r=r_0,\vartheta=\frac{\rpi}{2}}+\dots}{}
This important distinction seems to have been noticed only lately \cite{Zasl16} (for the $\tilde q=0$ case), although note it can be deduced from earlier calculations presented in \cite{Zasl11c}. There it is shown that critical particles have, unlike usual ones, a non-divergent redshift factor with respect to the stationary tetrad in the horizon limit. Therefore, although the BSW-type effects are often advertised as particle acceleration, they are in fact caused by \qt{slowness} of the critical particles.

Let us illustrate this in yet another way. We have already noticed that $\varepsilon=V$ at $r_0$ for critical particles, which implies $W=0$. Furthermore, it follows easily from \eqref{sqwcr} (or \eqref{wveffder}) that
\rov{\res{\pder{W}{r}}{r=r_0,\msc X_\mrm{H}=0}=0\rvt}{weffderzero}
Concurrence of these conditions would seem to suggest that there is a circular orbit at $r_0$ for parameters of each critical particle. However, there exist doubts about the properties of orbits in the $r=r_0$ region (cf. \cite{BarPrTeu, Jacobson11}). Regardless of these doubts, let us select an arbitrary radius $r_\mrm{orb}\geqslant r_+$, and see what it implies if we assume that $W$ and its first derivative are zero there. Expanding \eqref{urgen} around $r_\mrm{orb}$ (for $r\geqslant r_\mrm{orb}$), we get 
\rov{u^r\equiv\der{r}{\tau}\doteq\pm\(r-r_\mrm{orb}\)\res{\sqrt{\frac{1}{2{\tilde N}^2\tilde g_{rr}}\pder[2]{W}{r}}}{r=r_\mrm{orb}}+\dots}{}
This equation has an asymptotic solution of the form
\prov{r&\doteq r_\mrm{orb}\[1+\exp\(\pm\frac{\tau}{\tau_\mrm{relax}}\)\]+\dots&\frac{1}{\tau_\mrm{relax}}&=\res{\sqrt{\frac{1}{2{\tilde N}^2\tilde g_{rr}}\pder[2]{W}{r}}}{r=r_\mrm{orb}}\rvc}
which is valid for early proper times for outgoing particles (plus sign) and for late ones for incoming particles (minus sign). We can apply the result to critical particles by choosing the minus sign and $r_\mrm{orb}=r_0$. The result that critical particles only asymptotically approach the radius of the degenerate horizon and do not reach it in a finite proper time has already been derived in a slightly different way in various cases, see e.g. \cite{GP10a, Zasl10}. Above, we have shown that it applies to the $\tilde q\neq0,\omega\neq0$ case as well.

Let us yet mention that it follows from \eqref{sqwcr} (or \eqref{wveffdder}) that
\rov{\res{\pder[2]{W}{r}}{r=r_0,\msc X_\mrm{H}=0}=2\res{\(\pder{V_+}{r}\pder{V_-}{r}\)}{r=r_0}\rvt}{wveffdderhor}
This interconnection between derivatives of $W$ and $V_\pm$ of different orders appears rather unusual regarding the general form of \eqref{wveffdder}.

\subsection{The hyperbola}

As we have seen above, whether the motion of critical particles towards $r_0$ is forbidden or not depends on whether the radial derivative of $V$ at $r=r_0$ is positive or not. To study the division between the critical particles that can approach $r_0$ and those that cannot, we thus consider the condition
\rov{\res{\pder{V}{r}}{r=r_0}=0}{derveffzero}
as a function of parameters $\tilde q$ and $l$ of the (critical) test particles. Regarding \eqref{veffderh}, one sees that it actually corresponds to a branch of a hyperbola in variables $\tilde q$ and $l$. To study its properties, we remove the square root in \eqref{veffderh} and thus recover the second branch 
\rov{\res{\[\frac{\(\pder{\omega}{r}l+\tilde q\pder{\phi}{r}\)^2}{{\tilde N}^2}-\frac{\(l-\tilde qA_\varphi\)^2}{g_{\varphi\varphi}}\]}{r=r_0,\vartheta=\frac{\rpi}{2}}=1\rvt}{admhyp}
This equation corresponds in fact to
\rov{\res{\pder[2]{W}{r}}{\msc X_\mrm{H}=0,r=r_0}=0\rvt}{weffdderzero}
Regarding \eqref{wveffdderhor} (cf. also the calculations leading to \eqref{veffderh}), we find that the second branch corresponds to similar division for non-physical particles moving backwards in time.

We already observed that conditions \eqref{weffderzero} and $W=0$ always hold for critical particles at $r_0$: thus, the simultaneous validity of \eqref{weffdderzero} signifies the usual requirement for a marginally stable circular orbit, often called innermost stable circular orbit (ISCO). Then, from \eqref{wveffdderhor} we see that the condition \eqref{derveffzero} also implies the ISCO in this sense (restricting to particles moving forward in time it is in fact equivalent to the requirement for ISCO). 

The curve described by equation \eqref{admhyp} is indeed a hyperbola, except for the case when the two squared expressions become proportional to each other. This would happen if
\rov{\res{\(\pder{\phi}{r}+\pder{\omega}{r}A_\varphi\)}{r=r_0,\vartheta=\frac{\rpi}{2}}=0\rvt}{lindeg}
Calculating the derivative of the generalised electrostatic potential \eqref{aaxst},
\rov{\pder{\phi}{r}=-\pder{A_t}{r}-\omega\pder{A_\varphi}{r}-\pder{\omega}{r}A_\varphi\rvc}{elstgender}
and using the expression for the radial electric field strength in the locally non-rotating frame \eqref{lnrf1}, \eqref{lnrf2}, which reads
\rov{F_{(r)(t)}=\frac{1}{N\sqrt{g_{rr}}}\(\pder{A_t}{r}+\omega\pder{A_\varphi}{r}\)\rvc}{elstgenderalt}
we can write
\rov{\pder{\phi}{r}=-N\sqrt{g_{rr}}F_{(r)(t)}-\pder{\omega}{r}A_\varphi=-\tilde N\sqrt{\tilde g_{rr}}F_{(r)(t)}-\pder{\omega}{r}A_\varphi\rvt}{}
Since the product $N^2g_{rr}$ is finite and non-vanishing (as manifested by passing to $\tilde N^2$ and $\tilde g_{rr}$ in extremal case, cf. \eqref{axstex}), the condition \eqref{lindeg} reduces to
\rov{\res{F_{(r)(t)}}{r=r_0,\vartheta=\frac{\rpi}{2}}=0\rvt}{}
In this degenerate case equation \eqref{admhyp} defines just a pair of straight lines in $l\tilde q$ plane rather than a hyperbola.

The hyperbola \eqref{admhyp} has asymptotes
\rov{l=-\tilde q\res{\frac{\sqrt{g_{\varphi\varphi}}\pder{\phi}{r}\pm\tilde NA_\varphi}{\sqrt{g_{\varphi\varphi}}\pder{\omega}{r}\mp\tilde N}}{r=r_0,\vartheta=\frac{\rpi}{2}}\rvt}{admhypas}
We can also rewrite \eqref{admhyp} as
\rov{\res{\left\{l^2\[\frac{\(\pder{\omega}{r}\)^2}{{\tilde N}^2}-\frac{1}{g_{\varphi\varphi}}\]+\tilde q^2\[\frac{\(\pder{\phi}{r}\)^2}{{\tilde N}^2}-\frac{A_\varphi^2}{g_{\varphi\varphi}}\]+2l\tilde q\(\frac{\pder{\omega}{r}\pder{\phi}{r}}{{\tilde N}^2}+\frac{A_\varphi}{g_{\varphi\varphi}}\)\ri\}}{r=r_0,\vartheta=\frac{\rpi}{2}}=1\rvt}{admhyp1}
The coefficients in this form determine the orientation of the hyperbola with respect to axes $\tilde q$ and $l$. We will distinguish several cases and denote them by numbers and letters (by which they are identified in figures in Section \ref{sek:kn}).

Case {\bf 1a}: If 
\rov{\res{\[g_{\varphi\varphi}\(\pder{\omega}{r}\)^2-{\tilde N}^2\]}{r=r_0,\vartheta=\frac{\rpi}{2}}>0\rvc}{l2pos}
\eqref{admhyp1} is valid for $\tilde q=0$, which means that both branches exist for both signs of $\tilde q$, i.e. both cross the $l$ axis. 

Case {\bf 1b}: If, on the other hand, 
\rov{\res{\[g_{\varphi\varphi}\(\pder{\omega}{r}\)^2-{\tilde N}^2\]}{r=r_0,\vartheta=\frac{\rpi}{2}}<0\rvc}{l2neg}
\eqref{admhyp1} cannot be satisfied for $\tilde q=0$, so the branches are separated by the $l$ axis and each of them corresponds to different sign of $\tilde q$. The marginal case {\bf 1c} occurs when in \eqref{admhyp1}
\rov{\res{\[g_{\varphi\varphi}\(\pder{\omega}{r}\)^2-{\tilde N}^2\]}{r=r_0,\vartheta=\frac{\rpi}{2}}=0\rvc}{l2deg}
i.e. the coefficient multiplying $l^2$ vanishes. Comparing with \eqref{admhypas}, we see that this corresponds to one of the asymptotes having infinite slope, thus coinciding with the $l$ axis. 

A similar discussion applies to the coefficient of $\tilde q^2$.

Case {\bf 2a}: If 
\rov{\res{\[g_{\varphi\varphi}\(\pder{\phi}{r}\)^2-{\tilde N}^2A_\varphi^2\]}{r=r_0,\vartheta=\frac{\rpi}{2}}>0\rvc}{q2pos}
both branches of the hyperbola cross the $\tilde q$ axis and exist for both signs of $l$. 

Case {\bf 2b}: The opposite inequality,
\rov{\res{\[g_{\varphi\varphi}\(\pder{\phi}{r}\)^2-{\tilde N}^2A_\varphi^2\]}{r=r_0,\vartheta=\frac{\rpi}{2}}<0\rvc}{q2neg}
 means that the branches are separated by the $\tilde q$ axis and each of them has different sign of $l$. 
 
Case {\bf 2c}: If
\rov{\res{\[g_{\varphi\varphi}\(\pder{\phi}{r}\)^2-{\tilde N}^2A_\varphi^2\]}{r=r_0,\vartheta=\frac{\rpi}{2}}=0\rvc}{q2deg}
one can see again from \eqref{admhypas}  that this means that (at least) one of the asymptotes has zero slope, i.e. it coincides with the $\tilde q$ axis.

Case {\bf 3}: If 
\rov{\res{\(g_{\varphi\varphi}\pder{\omega}{r}\pder{\phi}{r}+{\tilde N}^2A_\varphi\)}{r=r_0,\vartheta=\frac{\rpi}{2}}=0\rvc}{lqdeg}
the coefficient multiplying $l\tilde q$ vanishes. In that case, the hyperbola is symmetrical with respect to the inversions $l\to-l$ and $\tilde q\to-\tilde q$. Since we are interested in just one of the branches, only one of the symmetries matters.

Let us note that when the electromagnetic field vanishes, the conditions \eqref{lindeg}, \eqref{q2deg} and \eqref{lqdeg} are satisfied simultaneously -- the charge of the particle loses any effect on kinematics.

Turning back to \eqref{veffderh} in general, it is obvious that the radial derivative of the effective potential at $r_0$ will be positive for $l=0, \tilde q=0$. Therefore, the \qt{admissible region} in the $l\tilde q$ plane will be \qt{outside} the hyperbola branch given by \eqref{derveffzero} with \eqref{veffderh}. If the branches are separated by one of the axes, critical particles must have a specific sign of one of the parameters to possibly reach $r_0$. Therefore, the difference among the {\bf a} and {\bf b} subcases of {\bf 1} and {\bf 2} is essential. 

It is even more important to look at combinations of these subcases. There are two generic possibilities, when one of the two BSW mechanisms prevails: the combination {\bf 1a2b}, when only the sign of $l$ is restricted, corresponds to the \qt{classical} centrifugal mechanism of BSW effect (first described in \cite{BSW} and generalised in \cite{Zasl10}). On the other hand, the variant {\bf 1b2a} with a restriction on the sign of $\tilde q$ signifies the dominance of the electrostatic analogy of the BSW effect (conceived in \cite{Zasl11a}). 

However, in the $\omega\neq0, \tilde q\neq0$ case, another two (more unusual) combinations can possibly occur. Scenario {\bf 1a2a} means that the  sign of neither $l$ nor $\tilde q$ is restricted. In this case, there will be critical particles with both signs of $l$ and with both signs of $\tilde q$ that can approach $r_0$. Just \emph{one} combination of signs of both parameters will be excluded. In contrast, the possibility {\bf 1b2b} would mean that signs of both $l$ and $\tilde q$ are restricted, i.e. that only critical particles with just one combination of signs of $l$ and $\tilde q$ can approach $r_0$. Curiously enough, for the extremal Kerr-Newman solution (see below), of those two, only the {\bf 1a2a} case can occur. However, the {\bf 1b2b} variant could possibly be realised in more general black-hole models. 

As the {\bf c} cases represent transitions between different combinations described above, the corresponding conditions \eqref{l2deg} and \eqref{q2deg} have particular physical significance; it is of primary interest, if these conditions can be satisfied for some black-hole solution and for which values of its parameters.

Finally, let us note that the condition \eqref{critgen} for critical particles can be used to define a system of parallel lines (labeled by different values of $\varepsilon_\mrm{cr}$) in the $l\tilde q$ plane.\footnote{Alternatively, one can also interpret \eqref{critgen} as an equation of single plane in $\varepsilon l\tilde q$ space, as we did in \cite{dipl}.} Apart from the orientation of the hyperbola, it is also of interest to examine how the branch defined by \eqref{derveffzero} with \eqref{veffderh} intersects these critical energy lines and which critical energies belong to the the admissible region.

\subsection{Parametric solution}

There are many possible parametrisations for branch(es) of a hyperbola. We will derive a particular parametrisation of the hyperbola branch given by \eqref{derveffzero} with \eqref{veffderh}, which is simple and which can also be used to describe curves given by analogues of \eqref{derveffzero} with higher derivatives.\footnote{One can also express the hyperbola as a pair of functions $l\f(\tilde q\)$ (as we examined in \cite{dipl}).}

Namely, let us make the following change of variables (assuming $\res{A_\varphi}{r=r_0,\vartheta=\nicefrac{\rpi}{2}}\neq0$):
\prov{l&=\lambda\eta\res{A_\varphi}{r=r_0,\vartheta=\frac{\rpi}{2}}\rvc&\tilde q&=\lambda\(1+\eta\)\rvc\lbl{lameta}}
under which \eqref{derveffzero} with \eqref{veffderh} becomes
\rov{\res{\(\pder{\omega}{r}\lambda\eta A_\varphi+\lambda\(1+\eta\)\pder{\phi}{r}+\tilde N\sqrt{1+\frac{\lambda^2A_\varphi^2}{g_{\varphi\varphi}}}\)}{r=r_0, \vartheta=\frac{\rpi}{2}}=0\rvt}{}
Expressing $\eta$ as
\rov{\eta=-\res{\frac{\lambda\pder{\phi}{r}+\tilde N\sqrt{1+\frac{\lambda^2A_\varphi^2}{g_{\varphi\varphi}}}}{\lambda\(\pder{\phi}{r}+\pder{\omega}{r}A_\varphi\)}}{r=r_0,\vartheta=\frac{\rpi}{2}}=\res{\frac{\lambda\pder{\phi}{r}+\tilde N\sqrt{1+\frac{\lambda^2A_\varphi^2}{g_{\varphi\varphi}}}}{\lambda\tilde N\sqrt{\tilde g_{rr}}F_{(r)(t)}}}{r=r_0,\vartheta=\frac{\rpi}{2}}\rvc}{}
and plugging back into \eqref{lameta}, we get parametric equations for $l, \tilde q$ as functions of $\lambda$ 
\rov{l=-\res{\(\frac{\lambda\pder{\phi}{r}+\tilde N\sqrt{1+\frac{\lambda^2A_\varphi^2}{g_{\varphi\varphi}}}}{\pder{\phi}{r}+\pder{\omega}{r}A_\varphi}A_\varphi\)}{r=r_0,\vartheta=\frac{\rpi}{2}}=\res{\(\frac{\lambda\pder{\phi}{r}+\tilde N\sqrt{1+\frac{\lambda^2A_\varphi^2}{g_{\varphi\varphi}}}}{\tilde N\sqrt{\tilde g_{rr}}F_{(r)(t)}}A_\varphi\)}{r=r_0,\vartheta=\frac{\rpi}{2}}\rvc}{lparam}
and
\rov{\tilde q=\res{\frac{\lambda\pder{\omega}{r}A_\varphi-\tilde N\sqrt{1+\frac{\lambda^2A_\varphi^2}{g_{\varphi\varphi}}}}{\pder{\phi}{r}+\pder{\omega}{r}A_\varphi}}{r=r_0,\vartheta=\frac{\rpi}{2}}=\res{\frac{-\lambda\pder{\omega}{r}A_\varphi+\tilde N\sqrt{1+\frac{\lambda^2A_\varphi^2}{g_{\varphi\varphi}}}}{\tilde N\sqrt{\tilde g_{rr}}F_{(r)(t)}}}{r=r_0,\vartheta=\frac{\rpi}{2}}\rvt}{qparam}

If $\res{A_\varphi}{r=r_0,\vartheta=\nicefrac{\rpi}{2}}=0$, \eqref{derveffzero} with \eqref{veffderh} is just linear in $\tilde q$, so we can solve for it directly, 
\rov{\tilde q=-\res{\frac{\pder{\omega}{r}l+\tilde N\sqrt{1+\frac{l^2}{g_{\varphi\varphi}}}}{\pder{\phi}{r}}}{r=r_0,\vartheta=\frac{\rpi}{2}}\rvt}{qdirect}

\subsection{The second derivative}

In order for the critical particles to approach $r=r_0$, their parameters $\tilde q, l$ must lie in the admissible region. This region in the $\tilde ql$ plane is delimited by the hyperbola branch given by the requirement of zero first derivative of the effective potential $V$ at $r_0$.\footnote{Let us note that in Table I of \cite{HK11b}, Harada and Kimura propose a classification (for nonequatorial critical particles in the Kerr field) somewhat similar to our discussion based on $\res{\nicepder{V}{r}}{r_0}$. In particular class I critical particles correspond to those inside our admissible region, class II to those on the border and class III to the ones outside of it.} There are, however, more subtle aspects. First, for critical particles with parameters located (almost) at the border of the admissible region, i.e. parameters corresponding to (almost) zero first derivative of $V$ at $r=r_0$, the second derivative of $V$ determines the trend of the effective potential $V$ and the admissibility of motion.

Furthermore, one should distinguish between the conditions suitable for \qt{black hole particle supercollider experiment}, where the motion towards $r=r_0$ should be allowed to start from some radius well above $r_0$ (if not from infinity of the spacetime, like in \cite{BSW}), and a situation, when the allowed band outside $r_0$    is tiny.\footnote{These other cases may be compatible with the particle starting to plunge after moving on a non-geodesic trajectory due to viscous losses inside an accretion disc. Such a process was discussed by Harada and Kimura in a slightly different context \cite{HK11a}.} If we do not make assumptions about asymptotics of the effective potential $V$ (or of the spacetime itself), this distinction also depends on the second derivative of $V$ at $r=r_0$. There are multiple possibilities inside the admissible region. If the first derivative of $V$ at $r=r_0$ is negative but very small, whereas the second one is positive, $V$ will reach a minimum and start to increase for some radii not much higher than $r_0$. Thus, the motion of the corresponding critical particle will be allowed only in a modest range of $r$. On the other hand, if both the first and the second derivative of $V$ at $r=r_0$ are negative, they will not be outweighed by higher Taylor orders until radii of multiples of $r_0$, so the motion can start well outside of the black hole. (See the Kerr-Newman example below, cf. Figure \ref{f:kn60dveff}.) The higher derivatives can make a difference, even at radii very close to $r_0$, only in the (uncommon) case when both the first and the second derivative of $V$ at $r=r_0$ will be very small. 

Focusing here on the second derivative of $V$ at $r=r_0$ for an extremal black hole \eqref{axstex}, let us proceed analogously to what we described for the first derivative, namely, observing $\nicedpder{N}{r}=2\nicepder{\tilde N}{r}+\(r-r_0\)\nicedpder{\tilde N}{r}$. The result is
\drov{\res{\dpder{V}{r}}{r=r_0}={}&\vast[\dpder{\omega}{r}l+\tilde q\dpder{\phi}{r}+\(2\pder{\tilde N}{r}-\frac{\tilde N}{g_{\varphi\varphi}}\pder{g_{\varphi\varphi}}{r}\)\sqrt{1+\frac{\(l-\tilde qA_\varphi\)^2}{g_{\varphi\varphi}}}+\zrov&+\res{\frac{\tilde N}{g_{\varphi\varphi}}\(\pder{g_{\varphi\varphi}}{r}-2\tilde q\(l-\tilde qA_\varphi\)\pder{A_\varphi}{r}\)\frac{1}{\sqrt{1+\frac{\(l-\tilde qA_\varphi\)^2}{g_{\varphi\varphi}}}}\vast]}{r=r_0,\vartheta=\frac{\rpi}{2}}\rvt}{veffdderh}
Again, we will consider the condition
\rov{\res{\dpder{V}{r}}{r=r_0}=0}{dderveffzero}
as a prescription of a curve in variables $\tilde q$ and $l$. First, one can deduce its asymptotes,
\rov{l=-\tilde q\res{\frac{\dpder{\phi}{r}\pm\frac{1}{\sqrt{g_{\varphi\varphi}}}\[\(2\pder{\tilde N}{r}-\frac{\tilde N}{g_{\varphi\varphi}}\pder{g_{\varphi\varphi}}{r}\)A_\varphi+2\tilde N\pder{A_\varphi}{r}\]}{\dpder{\omega}{r}\mp\frac{1}{\sqrt{g_{\varphi\varphi}}}\(2\pder{\tilde N}{r}-\frac{\tilde N}{g_{\varphi\varphi}}\pder{g_{\varphi\varphi}}{r}\)}}{r=r_0,\vartheta=\frac{\pi}{2}}\rvt}{dderas}
However, since \eqref{dderveffzero} leads to much more complicated curve than a branch of a hyperbola, the asymptotes do not provide good information. (In fact, for the Kerr-Newman solution it can be seen that in some cases the curve approaches the asymptotes very slowly and that it may also cross them.) 

Nevertheless, we can use change of variables \eqref{lameta} to obtain a parametric solution in the form 
\rov{l=-\res{\[\frac{\lambda\dpder{\phi}{r}\sqrt{1+\frac{\lambda^2A_\varphi^2}{g_{\varphi\varphi}}}+2\pder{\tilde N}{r}+\(2\pder{\tilde N}{r}-\frac{\tilde N}{g_{\varphi\varphi}}\pder{g_{\varphi\varphi}}{r}+2\frac{\tilde N}{A_\varphi}\pder{A_\varphi}{r}\)\frac{\lambda^2A_\varphi^2}{g_{\varphi\varphi}}}{\(\dpder{\omega}{r}A_\varphi+\dpder{\phi}{r}\)\sqrt{1+\frac{\lambda^2A_\varphi^2}{g_{\varphi\varphi}}}+2\tilde N\frac{\lambda A_\varphi}{g_{\varphi\varphi}}\pder{A_\varphi}{r}}A_\varphi\]}{r=r_0,\vartheta=\frac{\rpi}{2}}}{l2param}
and
\rov{\tilde q=\res{\frac{\lambda\dpder{\omega}{r}A_\varphi\sqrt{1+\frac{\lambda^2A_\varphi^2}{g_{\varphi\varphi}}}-2\pder{\tilde N}{r}-\(2\pder{\tilde N}{r}-\frac{\tilde N}{g_{\varphi\varphi}}\pder{g_{\varphi\varphi}}{r}\)\frac{\lambda^2A_\varphi^2}{g_{\varphi\varphi}}}{\(\dpder{\omega}{r}A_\varphi+\dpder{\phi}{r}\)\sqrt{1+\frac{\lambda^2A_\varphi^2}{g_{\varphi\varphi}}}+2\tilde N\frac{\lambda A_\varphi}{g_{\varphi\varphi}}\pder{A_\varphi}{r}}}{r=r_0,\vartheta=\frac{\rpi}{2}}\rvt}{q2param}
Once more, when $\res{A_\varphi}{r=r_0,\vartheta=\nicefrac{\rpi}{2}}=0$ and \eqref{lameta} does not work, we can solve \eqref{dderveffzero} directly for $\tilde q$, obtaining
\rov{\tilde q=-\res{\frac{\dpder{\omega}{r}l\sqrt{1+\frac{l^2}{g_{\varphi\varphi}}}+2\pder{\tilde N}{r}+\(2\pder{\tilde N}{r}-\frac{\tilde N}{g_{\varphi\varphi}}\pder{g_{\varphi\varphi}}{r}\)\frac{l^2}{g_{\varphi\varphi}}}{\dpder{\phi}{r}\sqrt{1+\frac{l^2}{g_{\varphi\varphi}}}-\frac{2\tilde N}{g_{\varphi\varphi}}l\pder{A_\varphi}{r}}}{r=r_0,\vartheta=\frac{\rpi}{2}}\rvt}{q2direct}
The details of the behaviour of the curve \eqref{dderveffzero} are in general not quite simple, since the curve can have two branches separated by a \qt{third asymptote}. This is manifested by the fact that denominators of \eqref{l2param} and \eqref{q2param} can go to zero for a finite value of parameter $\lambda$. One can verify that for real $\lambda$ the only such value can be
\rov{\lambda_0=-\res{\[\sgn\(\pder{A_\varphi}{r}\)\frac{\sqrt{g_{\varphi\varphi}}}{A_\varphi}\frac{\dpder{\omega}{r}A_\varphi+\dpder{\phi}{r}}{\sqrt{\frac{4{\tilde N}^2}{g_{\varphi\varphi}}\(\pder{A_\varphi}{r}\)^2-\(\dpder{\omega}{r}A_\varphi+\dpder{\phi}{r}\)^2}}\]}{r=r_0,\vartheta=\frac{\rpi}{2}}\rvt}{lamzero}
Taking the limit $\lambda\to\lambda_0$ of the ratio of \eqref{l2param} and \eqref{q2param}, we find the third asymptote to be the line $l=\tilde q\res{A_\varphi}{r=r_0,\vartheta=\nicefrac{\rpi}{2}}$. Since the values of $l$ and $\tilde q$ given by \eqref{l2param} and \eqref{q2param} for $\lambda=0$ lie on this line, we see that the curve \eqref{dderveffzero} necessarily crosses this third asymptote.

In order for the branch separation to occur, $\lambda_0$ must be a real number. Therefore, if it holds that
\rov{\res{\[\frac{4{\tilde N}^2}{g_{\varphi\varphi}}\(\pder{A_\varphi}{r}\)^2-\(\dpder{\omega}{r}A_\varphi+\dpder{\phi}{r}\)^2\]}{r=r_0,\vartheta=\frac{\rpi}{2}}>0\rvc}{}
the curve \eqref{dderveffzero} will indeed have two branches, whereas for
\rov{\res{\[\frac{4{\tilde N}^2}{g_{\varphi\varphi}}\(\pder{A_\varphi}{r}\)^2-\(\dpder{\omega}{r}A_\varphi+\dpder{\phi}{r}\)^2\]}{r=r_0,\vartheta=\frac{\rpi}{2}}<0\rvc}{2dsbr}
there will be only one branch. Interestingly, in the marginal case,
\rov{\res{\[\frac{4{\tilde N}^2}{g_{\varphi\varphi}}\(\pder{A_\varphi}{r}\)^2-\(\dpder{\omega}{r}A_\varphi+\dpder{\phi}{r}\)^2\]}{r=r_0,\vartheta=\frac{\rpi}{2}}=0\rvs}{2ddeg}
one can verify that the line $l=\tilde q\res{A_\varphi}{r=r_0,\vartheta=\nicefrac{\rpi}{2}}$ coincides with one of the asymptotes \eqref{dderas}.

The formulae \eqref{l2param} and \eqref{q2param} can be decomposed into two contributions
\prov{l\f(\lambda\)&=l_\mrm{reg}\f(\lambda\)+l_\mrm{sing}\f(\lambda\)\rvc&\tilde q\f(\lambda\)&=\tilde q_\mrm{reg}\f(\lambda\)+\tilde q_\mrm{sing}\f(\lambda\)\rvt\lbl{regsing}}
Here $l_\mrm{reg}$ and $\tilde q_\mrm{reg}$ are finite for $\lambda\to\lambda_0$, whereas $l_\mrm{sing}$ and $\tilde q_\mrm{sing}$ are given by expressions \eqref{l2param} and \eqref{q2param} with their numerators evaluated at $\lambda_0$. One can show that $l_\mrm{reg}, \tilde q_\mrm{reg}$ alone form a parametric expression of a branch of a hyperbola with asymptotes \eqref{dderas}, whereas $l_\mrm{sing}, \tilde q_\mrm{sing}$ parametrise the line $l=\tilde q\res{A_\varphi}{r=r_0,\vartheta=\nicefrac{\rpi}{2}}$. Unfortunately, the resulting expressions are not so \qt{practical} in general (see \ref{app:regsing}), but they become shorter for the Kerr-Newman case (cf. \eqref{l2regkn}, \eqref{q2regkn}, \eqref{lq2singkn}).

Leaving aside the technical details, let us note that it is of interest to study the intersections of curve \eqref{l2param}, \eqref{q2param} with the border \eqref{lparam}, \eqref{qparam} of the admissible region. If there is a part of the border that lies inside the region where the second derivative of $V$ at $r_0$ is positive, the cases described at the beginning of this section will arise. In the figures in the next section, these \qt{problematic} parts of the border will be plotted \textcolor{red}{in red}.

\section{Results for the Kerr-Newman solution}

\lbl{sek:kn}

For the Kerr-Newman solution with mass $M$, angular momentum $aM$ ($a\geqslant0$), and charge $Q$, the metric \eqref{axst} reads
\rov{\mbs g=-\frac{\iDelta\iSigma}{\msc A}\bd t^2+\frac{\msc A}{\iSigma}\sin^2\vartheta\[\bd\varphi-\frac{a}{\msc A}\(2Mr-Q^2\)\bd t\]^2+\frac{\iSigma}{\iDelta}\bd r^2+\iSigma\bd\vartheta^2\rvc}{kn}
where
\prov{\iDelta&=r^2-2Mr+a^2+Q^2\rvc&\iSigma&=r^2+a^2\cos^2\vartheta\rvc&\msc A&=\(r^2+a^2\)^2-\iDelta a^2\sin^2\vartheta\rvt}
In the extremal case, $M^2=Q^2+a^2$ and $\iDelta=\(r-M\)^2$, so $\iDelta$ plays the role of expression $\(r-r_0\)^2$ factored out in \eqref{axstex} with  $r_0=M$. It is obvious that one of the parameters, say $M$, constitutes just a scale; only ratios of the other two parameters with respect to it imply properties of the solution. Thus, the extremal case is effectively a one-parameter class.

The electromagnetic potential for the Kerr-Newman solution is
\rov{\mbs A=-\frac{Qr}{\iSigma}\(\bd t-a\sin^2\vartheta\bd\varphi\)\rvc}{akn}
which implies the generalised electrostatic potential,
\rov{\phi=\frac{Qr}{\msc A}\(r^2+a^2\)\rvt}{elstgenkn}

Substituting \eqref{kn}, \eqref{akn} and \eqref{elstgenkn} into \eqref{veffgen}, we get the effective potential for equatorial electrogeodesic motion (cf. \cite{Bardeen72})
\rov{V=\frac{1}{\msc A_\mrm{eq}}\left\{\(2Mr-Q^2\)al+\tilde qQr\(r^2+a^2\)+r\sqrt{\iDelta\[\msc A_\mrm{eq}+\(lr-\tilde qQa\)^2\]}\ri\}\rvc}{veffkn}
where $\msc A_\mrm{eq}$ stands for $\res{\msc A}{\vartheta=\nicefrac{\rpi}{2}}$. It is interesting to note that in the extremal case, for particles with special values of parameters 
\prov{l&=a\rvc&\tilde q&=\frac{\sqrt{Q^2+a^2}}{Q}\rvc\lbl{veffconst}}
it holds that $V\equiv1$. 

\subsection{The hyperbola}

Critical particles with given values of $\tilde q$ and  $l$ must have the energy defined by
\rov{\varepsilon_\mrm{cr}=\frac{al}{Q^2+2a^2}+\frac{\tilde qQ\sqrt{Q^2+a^2}}{Q^2+2a^2}\rvt}{critkn}
Kinematic restrictions on their motion towards $r=M$ are expressed by the branch of the hyperbola defined by equation \eqref{derveffzero} with \eqref{veffderh}, which for the extremal Kerr-Newman solution takes the form (when multiplied by common denominator $\(Q^2+2a^2\)^2$)
\rov{-2al\sqrt{Q^2+a^2}-\tilde qQ^3+\sqrt{Q^2+a^2}\sqrt{\(Q^2+2a^2\)^2+\(l\sqrt{Q^2+a^2}-\tilde qQa\)^2}=0\rvt}{derveffzerokn}
If we turn to the whole hyperbola in the form \eqref{admhyp}, we get
\rov{\frac{\(2al\sqrt{Q^2+a^2}+\tilde qQ^3\)^2}{\(Q^2+a^2\)\(Q^2+2a^2\)^2}-\frac{\(l\sqrt{Q^2+a^2}-\tilde qQa\)^2}{\(Q^2+2a^2\)^2}=1\rvt}{}
In the form \eqref{admhyp1}, it reads
\rov{l^2\frac{3a^2-Q^2}{\(Q^2+2a^2\)^2}+\tilde q^2\frac{Q^2\(Q^4-Q^2a^2-a^4\)}{\(Q^2+a^2\)\(Q^2+2a^2\)^2}+2l\tilde q\frac{Qa\(3Q^2+a^2\)}{\sqrt{Q^2+a^2}\(Q^2+2a^2\)^2}=1\rvt}{admhyp1kn}
Equation \eqref{admhypas} for the asymptotes of the hyperbola reduces to
\rov{l=\tilde q\frac{Q}{\sqrt{Q^2+a^2}}\frac{-Q^2\pm a\sqrt{Q^2+a^2}}{2a\pm\sqrt{Q^2+a^2}}\rvt}{admhypaskn} 

The parametric solution for \eqref{derveffzerokn} given in general by \eqref{lparam} and \eqref{qparam} turns out to be
\prov{l&=a\frac{-\lambda Q^3+\sqrt{Q^2+a^2}\sqrt{\(Q^2+2a^2\)^2+\lambda^2Q^2a^2}}{\sqrt{Q^2+a^2}\(Q^2+2a^2\)}\rvc\lbl{lparamkn}\\\tilde q&=\frac{2\lambda Qa^2+\sqrt{Q^2+a^2}\sqrt{\(Q^2+2a^2\)^2+\lambda^2Q^2a^2}}{Q\(Q^2+2a^2\)}\rvt\lbl{qparamkn}}

\subsection{The second derivative}

Regarding \eqref{veffdderh}, we find for the value of the second derivative of $V$ at $r=M$ 
\drov{\res{\dpder{V}{r}}{r=M}=\frac{1}{\(Q^2+2a^2\)^3}\Bigg[2a\(2Q^2+a^2\)l+2Q\sqrt{Q^2+a^2}\(Q^2-a^2\)\tilde q-\zrov-4Q^2\sqrt{\(Q^2+2a^2\)^2+\(l\sqrt{Q^2+a^2}-\tilde qQa\)^2}\Bigg]+\zrov+\frac{2}{\(Q^2+2a^2\)^2}\frac{Q^4+2Q^2a^2+\tilde qQa\(l\sqrt{Q^2+a^2}-\tilde qQa\)}{\sqrt{\(Q^2+2a^2\)^2+\(l\sqrt{Q^2+a^2}-\tilde qQa\)^2}}\rvt}{veffdderhkn}
One can check that for $\tilde q=0, l=0$ this expression reduces to
\rov{\res{\dpder{V}{r}}{r=M}=-\frac{2Q^2}{\(Q^2+2a^2\)^2}\rvc}{}
so that the region of $\res{\nicedpder{V}{r}}{r_0}>0$ will lie \qt{outside} the curve \eqref{dderveffzero}. The parametric equations \eqref{l2param} and \eqref{q2param} for this curve become 
\prov{l&=a\frac{-2Q^2a^2-Q^4+\frac{1}{Q^2+2a^2}\[\lambda Q\sqrt{Q^2+a^2}\(Q^2-a^2\)\sqrt{\(Q^2+2a^2\)^2+Q^2a^2\lambda^2}-\(3Q^2+2a^2\)Q^2a^2\lambda^2\]}{\lambda Qa^2\sqrt{Q^2+a^2}-Q^2\sqrt{\(Q^2+2a^2\)^2+Q^2a^2\lambda^2}}\rvc\lbl{l2paramkn}\\
\tilde q&=-\frac{Q^4+2Q^2a^2+\frac{1}{Q^2+2a^2}\[\lambda\frac{Qa^2}{\sqrt{Q^2+a^2}}\(2Q^2+a^2\)\sqrt{\(Q^2+2a^2\)^2+Q^2a^2\lambda^2}+2Q^4a^2\lambda^2\]}{\lambda Q^2a^2-\frac{Q^3}{\sqrt{Q^2+a^2}}\sqrt{\(Q^2+2a^2\)^2+Q^2a^2\lambda^2}}\rvt\lbl{q2paramkn}}

The value $\lambda_0$, for which the denominators of \eqref{l2paramkn} and \eqref{q2paramkn} vanish, is 
\rov{\lambda_0=\frac{Q}{a}\frac{Q^2+2a^2}{\sqrt{a^4+a^2Q^2-Q^4}}\rvt}{lamzerokn}

Performing the decomposition \eqref{regsing}, we find that finite part of \eqref{l2paramkn} is 
\rov{l_\mrm{reg}=a\frac{-Q\sqrt{Q^2+a^2}\(Q^4-4Q^2a^2-2a^4\)\lambda+\(2Q^4+2Q^2a^2+a^4\)\sqrt{\(Q^2+2a^2\)^2+Q^2a^2\lambda^2}}{\(Q^2+2a^2\)\(Q^4-Q^2a^2-a^4\)}\rvc}{l2regkn}
whereas for \eqref{q2paramkn} the finite part \eqref{q2reg} goes over to
\rov{\tilde q_\mrm{reg}=\frac{Qa^2\(4Q^4+3Q^2a^2\)\lambda+\(2Q^4+2Q^2a^2+a^4\)\sqrt{Q^2+a^2}\sqrt{\(Q^2+2a^2\)^2+Q^2a^2\lambda^2}}{Q\(Q^2+2a^2\)\(Q^4-Q^2a^2-a^4\)}\rvt}{q2regkn}
The contributions \eqref{l2regsing} and \eqref{q2sing} that blow up for $\lambda\to\lambda_0$ are given by\footnote{Interestingly, all the resulting formulae work well even for the case when real $\lambda_0$ does not exist \eqref{2dsbr}. They are not defined in the marginal case \eqref{2ddeg}.}
\prov{l_\mrm{sing}&=\frac{Qa\(Q^2+a^2\)}{Q^4-Q^2a^2-a^4}\frac{\(Q^2+2a^2\)^2}{\lambda a^2\sqrt{Q^2+a^2}-Q\sqrt{\(Q^2+2a^2\)^2+Q^2a^2\lambda^2}}\rvc&\tilde q_\mrm{sing}&=l_\mrm{sing}\frac{\sqrt{Q^2+a^2}}{Qa}\rvt\lbl{lq2singkn}}

The curves \eqref{lparamkn}, \eqref{qparamkn} and \eqref{l2paramkn}, \eqref{q2paramkn} have two intersections. Naturally, one of them coincides with \eqref{veffconst}. This corresponds to $V\equiv1$, as we stated before; thus, all the derivatives of $V$ at all radii will be zero for these values of $\tilde q$ and $l$. One can check that the point \eqref{veffconst} corresponds to $\lambda=0$ in \eqref{lparamkn}, \eqref{qparamkn} and \eqref{l2paramkn}, \eqref{q2paramkn}. The other intersection lies at
\prov{l&=\frac{a}{\left|Q\ri|}\frac{2Q^2+a^2}{\sqrt{Q^2+a^2}}\rvc&\tilde q&=\frac{1}{\left|Q\ri|}\frac{Q^2-a^2}{Q}\lbl{vtxlq}\rvc}
and corresponds to
\rov{\lambda=-\frac{1}{\left|Q\ri|}\frac{Q^2+2a^2}{Q}\rvt}{vtxlam}
One can make sure that the second derivative of $V$ at $r=M$ is positive on a finite stretch of the curve \eqref{lparamkn}, \eqref{qparamkn}, which lies in between these intersections. This part of the curve is plotted \textcolor{red}{in red} in the corresponding figures.

\subsection{Important special cases}

Let us now examine kinematic restrictions coming from the equations above for some specific cases of the extremal Kerr-Newman solution.

\subsubsection{Extremal Kerr solution}

The condition \eqref{lindeg} for the hyperbola branch \eqref{derveffzerokn} to degenerate into straight line corresponds to
\rov{\frac{Q}{Q^2+2a^2}=0}{}
for the extremal Kerr-Newman solution. We see that this can be satisfied only by setting $Q=0$, i.e. for the extremal Kerr solution. Regarding \eqref{admhyp1kn}, the conditions \eqref{q2deg} and \eqref{lqdeg} (case {\bf 2c3}) are also satisfied for $Q=0$, which we anticipated for a case without an electromagnetic field. 

Equation \eqref{derveffzerokn} is satisfied for $\nicefrac{l}{M}\equiv\nicefrac{l}{a}=\nicefrac{2}{\sqrt{3}}$. Therefore, critical particles can approach $r=M$ for angular momenta $\nicefrac{l}{M}>\nicefrac{2}{\sqrt{3}}$ in this case, which corresponds to energies $\varepsilon_\mrm{cr}>\nicefrac{1}{\sqrt{3}}$, as seen from \eqref{critkn}. However, the expression \eqref{veffdderhkn} becomes 
\rov{\res{\dpder{V}{r}}{r=M}=\frac{l}{4a^3}\rvc}{}
so the second derivative of $V$ at $r=M$ will be positive for all those particles. Thus, for the bounded particles close to $\nicefrac{l}{M}=\nicefrac{2}{\sqrt{3}}$, their motion will be allowed only for a short range of radii.
  
Let us yet note that the parameters $\nicefrac{l}{M}=\nicefrac{2}{\sqrt{3}},\varepsilon=\nicefrac{1}{\sqrt{3}}$ mentioned above are those of the marginally stable circular orbit in the extremal Kerr limit, as given in \cite{BarPrTeu}. Furthermore, the other special circular orbits considered in \cite{BarPrTeu}, i.e. the marginally bound orbit and the photon orbit (with $\varepsilon\to\infty, l\to2M\varepsilon$), also correspond to critical particles in the extremal Kerr limit.

\subsubsection{\qt{Sixty-degree} black hole}

Turning again to \eqref{admhyp1kn}, we find that condition \eqref{l2deg} (case {\bf 1c}) can be satisfied if and only if $3a^2=Q^2$. This corresponds to $\nicefrac{a}{M}=\nicefrac{1}{2}$ and $\nicefrac{\left|Q\ri|}{M}=\nicefrac{\sqrt{3}}{2}$, respectively.\footnote{If we express the parameters of the extremal Kerr-Newman black hole by a \qt{mixing angle} defined by $a=M\cos\gamma_\mrm{KN}, Q=M\sin\gamma_\mrm{KN}$, as we did in \cite{Article1}, $\nicefrac{a}{M}=\nicefrac{1}{2}$ corresponds to its value of sixty degrees, hence the name for this case.} The special alignment of the hyperbola branch \eqref{derveffzerokn} and the critical energy lines in the admissible region in this case can be seen in Figure \ref{f:exknps_1_2}.

In order to explore the significance of the sign of $\res{\nicedpder{V}{r}}{r_0}$, we plotted $V$ for several particles on the \qt{border} (with $\res{\nicepder{V}{r}}{r_0}$) in Figure \ref{f:kn60dveff}. With $\res{\nicedpder{V}{r}}{r_0}<0$, even the bounded particle shown (with $\tilde q=0.5$) has a reasonable allowed band.

\begin{figure}
\centering
\input{KN60DVeff.tex}
\caption{Effective potential $V$ for several particles with $\res{\nicepder{V}{r}}{r=r_0}=0$ moving around the extremal Kerr-Newman black hole with $\nicefrac{a}{M}=\nicefrac{1}{2}$. For $\nicefrac{2}{3}<\tilde q<\nicefrac{2}{\sqrt{3}}$ (\textcolor{red}{in red}), the effect of the positive second derivative of $V$ at $r_0$ is clearly visible.}
\label{f:kn60dveff}
\end{figure}
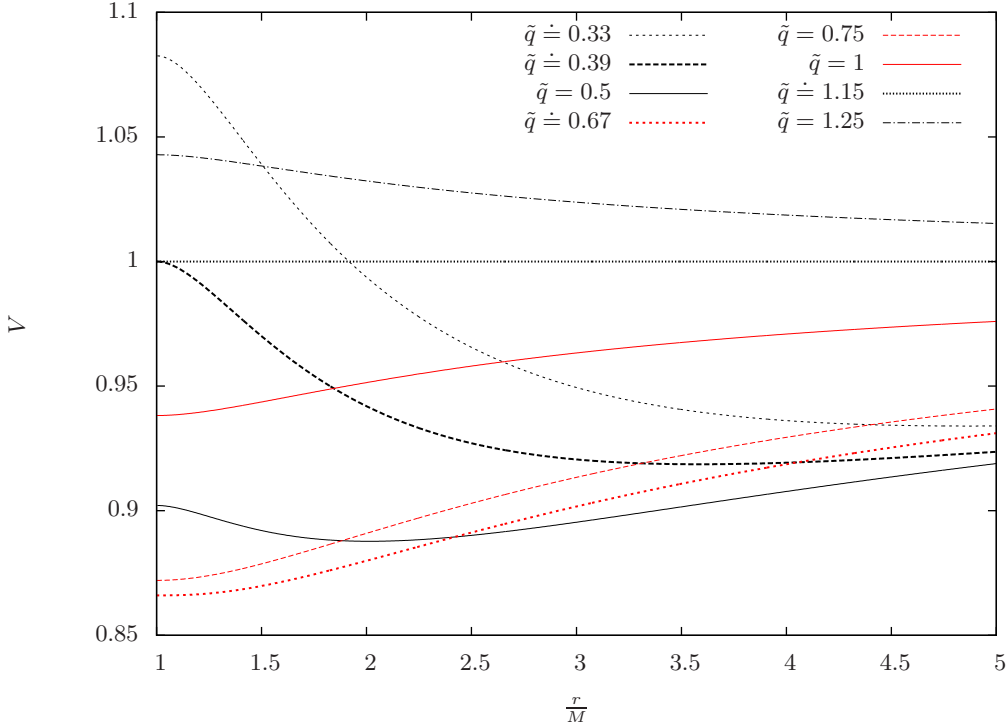

\begin{figure}
\centering
\input{exKNeqPS_1_2.tex}
\caption{Case {\bf 1c2a}: Kinematic restrictions for critical particles in the case of the extremal Kerr-Newman black hole with $\nicefrac{a}{M}=\nicefrac{1}{2}$. The hyperbola branch forms a border between the critical particles that can approach $r=M$ and those that cannot. In the admissible region the lines of constant critical energy are plotted. We considered $Q>0$; the figure for $Q<0$ can be obtained by the inversion $\tilde q\to-\tilde q$. Note that one of the asymptotes coincides with the $l$ axis.}
\label{f:exknps_1_2}
\end{figure}

\subsubsection{\qt{Golden} black hole}

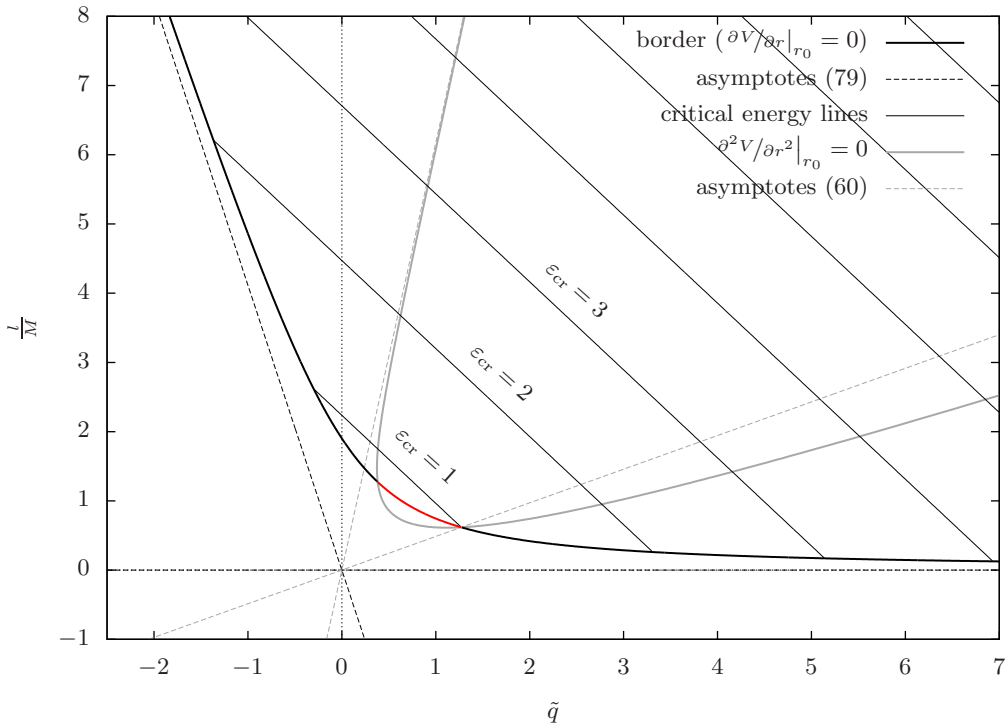
\begin{figure}
\centering
\input{exKNeqPS_s5-1_2.tex}
\caption{Case {\bf 1a2c}: Kinematic restrictions for critical particles in the case of the extremal Kerr-Newman black hole with $\nicefrac{a}{M}=\nicefrac{\(\sqrt{5}-1\)}{2}$ (golden black hole). The hyperbola branch forms a border between the critical particles that can approach $r=M$ and those that cannot. In the admissible region the lines of constant critical energy are plotted. We considered $Q>0$; the figure for $Q<0$ can be obtained by the inversion $\tilde q\to-\tilde q$. Note that one of the asymptotes coincides with the $\tilde q$ axis.}
\label{f:exknps_s5-1_2}
\end{figure}

Apart from the degenerate Kerr case, condition \eqref{q2deg} (case {\bf 2c}) will be also satisfied if $Q^4-Q^2a^2-a^4=0$, as follows from \eqref{admhyp1kn}. This equation has one positive root, which corresponds to $\nicefrac{Q^2}{a^2}=\nicefrac{\(\sqrt{5}+1\)}{2}$. This is the \qt{golden ratio} number.\footnote{F. H. thanks Miguel Coelho Ferreira for kindly pointing this out.} Since the golden ratio plus one equals the golden ratio squared, we get $\nicefrac{M}{a}=\nicefrac{\(\sqrt{5}+1\)}{2}$ and $\nicefrac{M^2}{Q^2}=\nicefrac{\(\sqrt{5}+1\)}{2}$. And, by definition, one over the golden ratio is the golden ratio minus one, so it holds that, e.g. $\nicefrac{a}{M}=\nicefrac{\(\sqrt{5}-1\)}{2}$. The plot of the hyperbola branch \eqref{derveffzerokn} in this case can be seen in Figure \ref{f:exknps_s5-1_2}.

Curiously enough, the condition \eqref{2ddeg} also corresponds to $Q^4-Q^2a^2-a^4=0$, as seen from \eqref{lamzerokn}. Thus, for $\nicefrac{a}{M}>\nicefrac{\(\sqrt{5}-1\)}{2}$ there will be two branches of the curve \eqref{l2paramkn}, \eqref{q2paramkn}. However, it turns out that only one of the branches will intersect the admissible region. This follows from the fact that one of the intersections with its border, curve \eqref{derveffzerokn}, lies at $\lambda=0$ and the position in $\lambda$ (cf. \eqref{vtxlam}) of the other has always the opposite sign compared to $\lambda_0$ (see \eqref{lamzerokn}), where the branch cut occurs.

We have exhausted all cases when conditions \eqref{l2deg} and \eqref{q2deg} can be satisfied. Now let us look at the signs of the coefficients multiplying $l^2$ and $\tilde q^2$ terms in \eqref{admhyp1kn} on the intervals delimited by these special cases. For $\nicefrac{a}{M}>\nicefrac{\(\sqrt{5}-1\)}{2}\doteq0.618$, the coefficient of $l^2$ is positive and the coefficient of $\tilde q^2$ is negative (case {\bf 1a2b}). Thus, the critical particles need to be corotating in order to approach $r=M$, but they can have both signs of charge (or be uncharged). The \qt{centrifugal mechanism} prevails in this interval. On the other hand, for $\nicefrac{a}{M}<\nicefrac{1}{2}$ the coefficient of $l^2$ is negative and the coefficient of $\tilde q^2$ is positive (case {\bf 1b2a}). Therefore, critical particles can be radially moving, counterrotating, or corotating, but they must have the same sign of charge as the black hole. In particular, they cannot be uncharged (cf. \cite{Liu13}). Thus, in this interval the \qt{electrostatic mechanism} prevails. However, for $\nicefrac{\(\sqrt{5}-1\)}{2}>\nicefrac{a}{M}>\nicefrac{1}{2}$, both coefficients are positive (case {\bf 1a2a}, so inequalities \eqref{l2pos} and \eqref{q2pos} hold simultaneously). In this interval one can choose between the mechanisms; the critical particles need to either be corotating or have the same sign of charge as the black hole in order to approach $r=M$.

\subsubsection{Extremal Reissner-Nordström solution}

\begin{figure}
\centering
\input{exKNeqPS_0.tex}
\caption{Case {\bf 1b2a3}: Kinematic restrictions for critical particles in the case of the extremal Reissner-Nordström black hole ($a=0$). The hyperbola branch forms a border between the critical particles that can approach $r=M$ and those that cannot. In the admissible region the lines of constant critical energy are plotted. We considered $Q>0$; the figure for $Q<0$ can be obtained by the inversion $\tilde q\to-\tilde q$. Note the symmetry with respect to the $\tilde q$ axis.}
\label{f:exknps_0}
\end{figure}
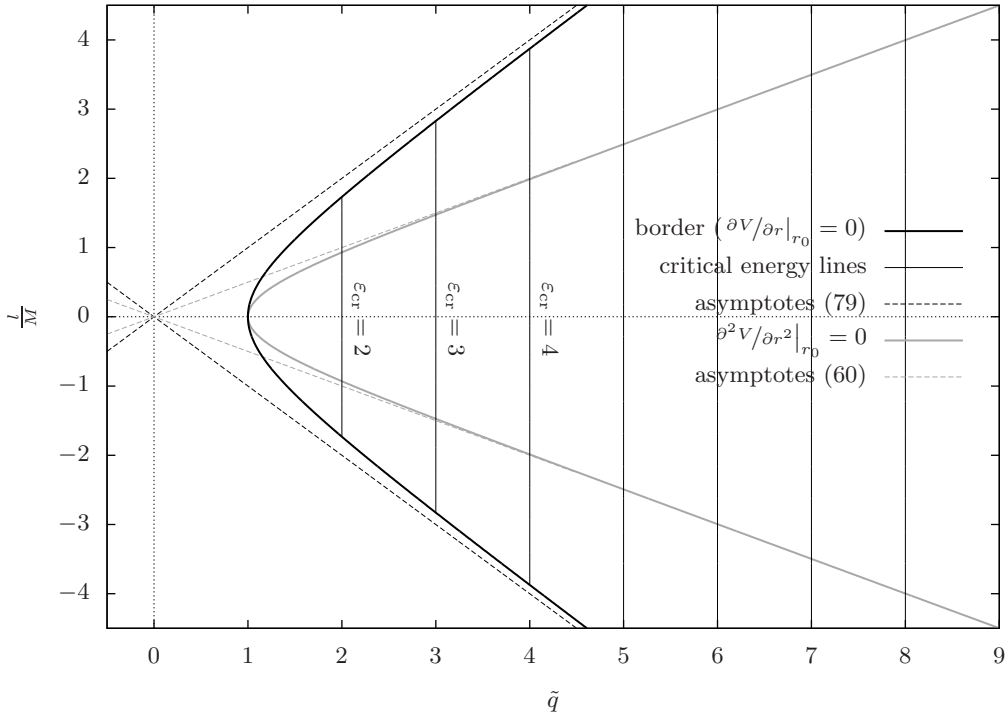

Again, in addition to the Kerr case, condition \eqref{lqdeg} (case {\bf 3}) can also be satisfied for $a=0$, as we see from \eqref{admhyp1kn}. As this case of the extremal Reissner-Nordström black hole is non-rotating, particle kinematics cannot depend on the change $l\to-l$. This is reflected in the symmetry of the hyperbola branch \eqref{derveffzerokn} with respect to $\tilde q$ axis, see Figure \ref{f:exknps_0}.

Since $A_\varphi\equiv0$ for $a=0$, one has to use solution \eqref{qdirect} for the hyperbola branch \eqref{derveffzerokn}, which reads
\rov{\tilde q=\frac{1}{Q}\sqrt{Q^2+l^2}\rvc}{}
and the solution \eqref{q2direct} for \eqref{dderveffzero}, which becomes
\rov{\tilde q=\frac{2l^2+Q^2}{Q\sqrt{Q^2+l^2}}\rvt}{}
These curves touch at $l=0, \tilde q=\sgn Q$. Let us note that radial critical particles were previously studied by Zaslavskii in \cite{Zasl11a}.

\subsection{Energy considerations}

As mentioned above, it is also of interest to study the intersections of the hyperbola branch \eqref{derveffzerokn} with critical energy lines. Since the Kerr-Newman solution is asymptotically flat, we focus on energy line with $\varepsilon_\mrm{cr}=1$, which corresponds to critical particles coming from rest at infinity. Solving for intersections of \eqref{critkn} for $\varepsilon_\mrm{cr}=1$ with \eqref{derveffzerokn}, we find that one is at the point \eqref{veffconst}, where $V\equiv1$ (all radial derivatives vanish for these parameters), and the other one occurs for
\prov{l&=a\frac{3Q^2+2a^2}{Q^2}\rvc&\tilde q&=\frac{\sqrt{Q^2+a^2}}{Q^3}\(Q^2-2a^2\)\rvt\lbl{eunint}}
Both intersections coincide for $a=0$, when they reduce to $l=0, \tilde q=\sgn Q$ and lie on the $\tilde q$ axis. Apart from this case, both intersections occur for positive $l$, so critical particles with $\varepsilon_\mrm{cr}\leqslant1$ must be always corotating for $a\neq0$. Only the second intersection can lie on the $l$ axis, which happens for $Q^2=2a^2$. This condition corresponds to $\nicefrac{a}{M}=\nicefrac{1}{\sqrt{3}}, \nicefrac{\left|Q\ri|}{M}=\sqrt{\nicefrac{2}{3}}$. Thus, we reproduced the result of \cite{WeiLiuGuoFu} that uncharged critical particles  with $\varepsilon_\mrm{cr}=1$ cannot approach $r=M$ for an extremal Kerr-Newman black hole with $\nicefrac{a}{M}<\nicefrac{1}{\sqrt{3}}$. The hyperbola branch \eqref{derveffzerokn} for this case is plotted in Figure \ref{f:exknps_1_s3}. For $Q\to0$, the expressions \eqref{veffconst} and \eqref{eunint} break down, because for the $Q=0$ (Kerr) case there is no dependence of the particle kinematics on $\tilde q$. In that case both \eqref{derveffzerokn} and \eqref{critkn} reduce just to (non-intersecting) lines of constant $l$.

\begin{figure}
\centering
\input{exKNeqPS_1_s3.tex}
\caption{Case {\bf 1a2a}: Kinematic restrictions for critical particles in the case of the extremal Kerr-Newman black hole with $\nicefrac{a}{M}=\nicefrac{1}{\sqrt{3}}$. The hyperbola branch forms a border between the critical particles that can approach $r=M$ and those that cannot. In the admissible region the lines of constant critical energy are plotted. We considered $Q>0$; the figure for $Q<0$ can be obtained by the inversion $\tilde q\to-\tilde q$. Note that the $\varepsilon_\mrm{cr}=1$ intersects with the border at the $l$ axis.}
\label{f:exknps_1_s3}
\end{figure}

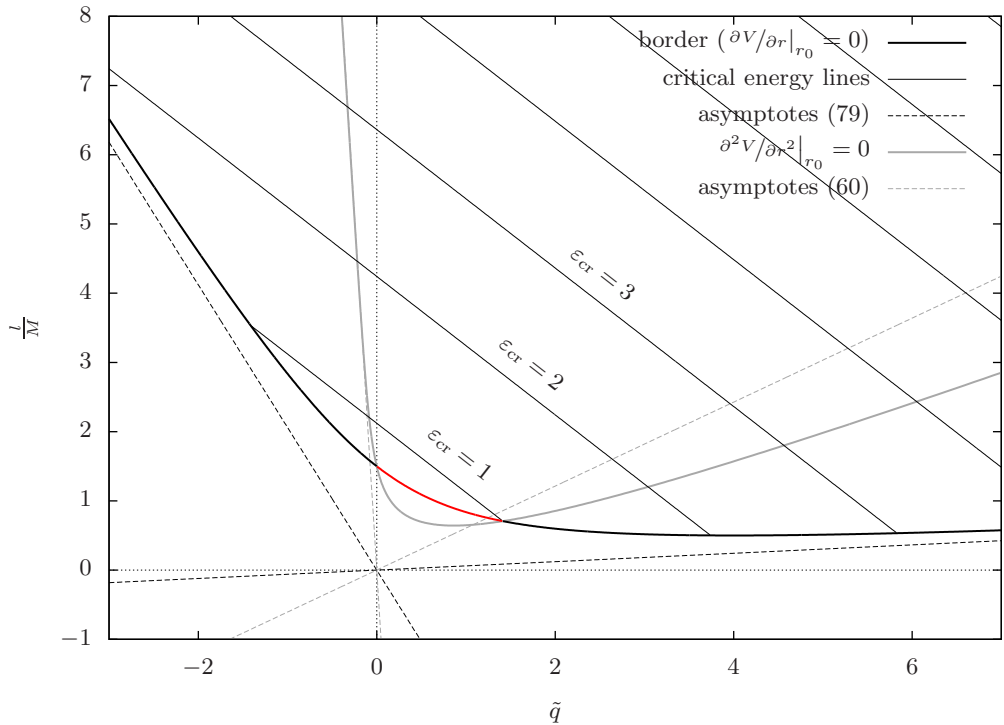
\begin{figure}
\centering
\input{exKNeqPS_1_s2.tex}
\caption{Case {\bf 1a2b}: Kinematic restrictions for critical particles in the case of the extremal Kerr-Newman black hole with $\nicefrac{a}{M}=\nicefrac{1}{\sqrt{2}}$. The hyperbola branch forms a border between the critical particles that can approach $r=M$ and those that cannot. In the admissible region the lines of constant critical energy are plotted. We considered $Q>0$; the figure for $Q<0$ can be obtained by the inversion $\tilde q\to-\tilde q$.}
\label{f:exknps_1_s2}
\end{figure}

Another interesting question is to find the \qt{energy vertex} of the hyperbola branch, i.e. what is the minimal value of the critical energy on curve \eqref{derveffzerokn}. One finds that it lies at the point \eqref{vtxlq}, with the corresponding critical energy being
\rov{\varepsilon_\mrm{cr}=\frac{\left|Q\ri|}{\sqrt{Q^2+a^2}}\rvt}{vtxen}
This vertex energy will always be smaller than 1, except for the $a=0$ (Reissner-Nordström) case, when the vertex coincides with the intersections with the $\varepsilon_\mrm{cr}=1$ line (\eqref{veffconst} and \eqref{eunint}) and lies on the $\tilde q$ axis. The vertex can cross the $l$ axis, which occurs if $Q^2=a^2$. That corresponds to $\nicefrac{a}{M}=\nicefrac{\left|Q\ri|}{M}=\nicefrac{1}{\sqrt{2}}$ (see Figure \ref{f:exknps_1_s2}).

Let us note that although the expressions \eqref{vtxlq} break down for $Q\to0$, the corresponding critical energy \eqref{vtxen} is regular. However, it does not have the correct limit for $Q\to0$, since it goes to zero, whereas the lowest energy required for critical particles in the extremal Kerr solution in order to approach $r=M$ is $\nicefrac{1}{\sqrt{3}}$, as noted above. This is an example of a \qt{discontinuous} behaviour of the kinematic restrictions in the limit $Q\to0$; it has further manifestations that we discuss below.

\subsection{The mega-BSW phenomena}

If $Q$ is small but nonzero, however tiny it may be, one can still maintain the magnitude of electrostatic force carried to a particular test particle if it has accordingly high $\tilde q$. This can be related to the divergent behaviour that we noticed in the expressions for positions of special points (all of them  corresponding to $\varepsilon_\mrm{cr}\leqslant1$) \eqref{veffconst}, \eqref{vtxlq} and \eqref{eunint} in the $l\tilde q$ plane. These features still occur regardless of how small the charge $Q$ is, but at higher and higher values of $\tilde q$. We can see that in all these cases it holds that $\left|l\ri|\doteq\left|Q\tilde q\ri|$ for very small $Q$. However, the divergences in the expressions are of different orders, which has interesting consequences.
   
Though the position of the intersection \eqref{veffconst} in $l$ approaches a constant for $Q\to0$ and just the position in $\tilde q$ diverges, for the other two points even the position in $l$ diverges for $Q\to0$. Thus, for very small $Q$, we can have charged critical particles that have $\varepsilon_\mrm{cr}\leqslant1$, yet posses enormous values of angular momentum, and which still can approach $r=M$ (hence the mega-BSW effect). Such a thing is not possible in either the $Q\approx a$ or $Q=0$ regimes.

To examine this effect in more detail, let us assume that there is some value $\tilde q_\mrm{max}\gg1$ that acts as an upper bound for specific charge of the particles, $\left|\tilde q\ri|\leqslant\tilde q_\mrm{max}$. Then we can find a value $Q_\mrm{min}$ of black hole's charge such that for $\left|Q\ri|\geqslant Q_\mrm{min}$ some of the special points (\eqref{veffconst}, \eqref{vtxlq}, or \eqref{eunint}) will fit in the interval $\[-\tilde q_\mrm{max},\tilde q_\mrm{max}\]$. The $Q\to0$ behaviour will be parametrised by $\tilde q_\mrm{max}\to\infty$ asymptotics. Furthermore, we can define $l_\mrm{max}$ such that the position of a selected special point (\eqref{veffconst}, \eqref{vtxlq}, or \eqref{eunint}) will be $\left|l\ri|\doteq l_\mrm{max}$ for $\left|Q\ri|=Q_\mrm{min}$. Since $Q_\mrm{min}$ will be small, we can use approximations and then observe the asymptotics for the three special points, which are summarised in Table \ref{tab:mega}.

\begin{table}
\caption{The mega-BSW effect illustrated (see text for details).}
\setstretch{1.4}
\begin{tabular}{l|c|c}
Point&$\nicefrac{Q_\mrm{min}}{M}$&$\nicefrac{l_\mrm{max}}{M}$\\
\hline
\eqref{veffconst}&$\(\tilde q_\mrm{max}\)^{-1}$&1\\
\hline
\eqref{vtxlq}&$\(\tilde q_\mrm{max}\)^{-\frac{1}{2}}$ &$\(\tilde q_\mrm{max}\)^\frac{1}{2}$\\
\hline
\eqref{eunint}&$2^\frac{1}{3}\(\tilde q_\mrm{max}\)^{-\frac{1}{3}}$ &$2^\frac{1}{3}\(\tilde q_\mrm{max}\)^\frac{2}{3}$
\end{tabular}
\lbl{tab:mega}
\end{table}

These asymptotics tell us how small is the value of $Q$, for which we can still fit one of the special points into the bounded range of values of charge $\tilde q$, and how large angular momentum $l$ the particles corresponding to this point can have for that value of $Q$.

This effect can also be relevant for considering BSW-type effects as an edge case for possible astrophysical particle collision processes. There are calculations, first using Wald's approximate (test-field) solution \cite{Wald74} and later an exact Ernst-Wild solution \cite{Hiscock81}, showing that a black hole can maintain a small, non-zero charge in the presence of an external magnetic field. Furthermore, considering elementary particles, an electron gives us $\tilde q_\mrm{max}>10^{20}$. However, the practical viability of the generalised BSW effect is in any case hindered by the unlikely existence of extremal black holes, the validity of the test-particle approximation and complications with energy extraction (cf. the Introduction).

\section{Summary and conclusion}

We have studied the kinematics of critical particles moving around axially symmetric stationary extremal black holes, focusing on the case when \emph{both} rotation and electromagnetic interaction are present. In the discussion, we used the minimum energy $V$ (equation \eqref{veffgen}), which is an analogy of a classical potential. Whether a critical particle can approach the position of the degenerate horizon or not depends heavily on properties of the black hole as well as on the parameters of the particle. If we treat the black hole as fixed, we can visualise the restrictions in the space of the parameters $l$ and $\tilde q$ (specific axial angular momentum and charge) of the particle. 

To do so, we derived expressions for curves $\res{\nicepder{V}{r}}{r_0}=0$, see \eqref{lparam}, \eqref{qparam}, and $\res{\nicedpder{V}{r}}{r_0}=0$, cf. \eqref{l2param}, \eqref{q2param}, in this parameter space. The first is just a branch of a hyperbola, whereas the second is technically complicated and can split into two branches. These curves divide the parameter space into different regions. Critical particles with parameters in the $\res{\nicepder{V}{r}}{r_0}<0$ part (the admissible region) can approach $r_0$. However, the interval of $r$ for which the motion is allowed may be short, if they fall into the part, where $\res{\nicedpder{V}{r}}{r_0}>0$.

We then studied the dependence of the restrictions on the properties of the black hole. The relevant question is how many quadrants of the $l\tilde q$ plane are intersected by the hyperbola branch (forming the border of the admissible region). As the admissible region is outside the hyperbola branch, it lies in the same quadrants as its border. In general cases it passes through two quadrants. In the case that we denoted as {\bf 1a2b}, critical particles must have a specific sign of angular momentum in order to approach $r=r_0$, but can have either sign of charge. Specially, they can be uncharged, but cannot move purely radially. This means the dominance of the centrifugal type of generalised BSW effect. On the other hand, in case {\bf 1b2a} the particles must have a specific sign of charge to approach $r=r_0$, but they can have either sign of angular momentum. This corresponds to the electrostatic type of generalised BSW effect. 

Furthermore, we found that two mixed cases are also possible. In case {\bf 1a2a}, the hyperbola branch passes through three quadrants, so that the signs of the charge and angular momentum of the critical particles are not restricted in order to approach $r=r_0$. Just one combination of the signs is forbidden. In contrast, in case {\bf 1b2b}, the signs of both charge and angular momentum of the critical particle approaching $r=r_0$ are restricted. We denoted the special limiting cases between {\bf a} and {\bf b} as {\bf c} (see also Table \ref{tab:sum}). 
Another special situation is case {\bf 3}, when the border (and therefore the whole admissible region) has the symmetry with respect to one of the inversions $l\to-l$ or $\tilde q\to-\tilde q$. The hyperbola branch may also degenerate into a straight line. We noted that this naturally happens for a vacuum black hole, together with conditions {\bf 2c3}.

We applied and illustrated the general discussion summarised above on the one-parameter class of extremal Kerr-Newman solutions. From the mixed cases, only {\bf 1a2a} is realised in this class. Apart from general kinematic restrictions (embodied in the position of the hyperbola branch enclosing the whole admissible region), we also investigated a subset of critical particles with energies corresponding to coming from rest at infinity or lower, $\varepsilon_\mrm{cr}\leqslant1$, i.e. marginally bound and bound particles. We found that for $\nicefrac{a}{M}>\nicefrac{1}{\sqrt{3}}$ these particles can have either sign of charge, but must be corotating to approach $r=M$, whereas for $\nicefrac{a}{M}<\nicefrac{1}{\sqrt{3}}$ they must be both corotating and have the same sign of the charge as the black hole in order to approach $r=M$. The restrictions for particles with $\varepsilon_\mrm{cr}\leqslant1$ are thus more stringent. The main results for restrictions on the parameters of critical particles in order to approach $r=M$ for extremal Kerr-Newman black holes are summarised in Table \ref{tab:sum}.\footnote{For convenience, apart from the ratio $\nicefrac{a}{M}$, we used also the Kerr-Newman mixing angle ($Q=M\sin\gamma_\mrm{KN}, a=M\cos\gamma_\mrm{KN}$) to parametrise the class.} 

\begin{table}

\caption{Restrictions on signs of $l,\tilde q$ for critical particles that can approach $r=M$ in an extremal Kerr-Newman spacetime. Special positions of special points in their parameter space are also indicated. (Note that \eqref{vtxlq} is the point with smallest $\varepsilon_\mrm{cr}$ in the admissible region, whereas \eqref{eunint} and \eqref{veffconst} are special points on its border, curve $\res{\nicepder{V}{r}}{r_0}=0$, with $\varepsilon_\mrm{cr}=1$. In the $a=0$ case, all three points coincide.)}
\setstretch{1.4}
\begin{tabular}{c|c|c|c|c}
\multicolumn{2}{c|}{Kerr-Newman black-hole parameters}&\multicolumn{2}{c|}{Restrictions}&Notes\\
$\left|\gamma_\mrm{KN}\ri|$& $\frac{a}{M}$&General case&For $\varepsilon_\mrm{cr}\leqslant1$\\
\hline
\hline
$0^\circ$&$1$&Vacuum&\multirow{6}{*}{$l>0$}\\
\cline{1-3}
$0^\circ<\left|\gamma_\mrm{KN}\ri|<45^\circ$&$1>\frac{a}{M}>\frac{1}{\sqrt{2}}$&\multirow{3}{*}{\bf 1a2b}&\\
\cline{5-5}
$45^\circ$&$\frac{1}{\sqrt{2}}$&&&\eqref{vtxlq} at $\tilde q=0$\\
\cline{5-5}
$45^\circ<\left|\gamma_\mrm{KN}\ri|<51.8^\circ$&$\frac{1}{\sqrt{2}}>\frac{a}{M}>\frac{\sqrt{5}-1}{2}$&&\\
\cline{1-3}
$\left|\gamma_\mrm{KN}\ri|\doteq51.8^\circ$&$\frac{\sqrt{5}-1}{2}$&{\bf 1a2c}&\\
\cline{1-3}
$51.8^\circ<\left|\gamma_\mrm{KN}\ri|<54.7^\circ$&$\frac{\sqrt{5}-1}{2}>\frac{a}{M}>\frac{1}{\sqrt{3}}$&\multirow{3}{*}{\bf 1a2a}&\\
\cline{4-5}
$\left|\gamma_\mrm{KN}\ri|\doteq54.7^\circ$&$\frac{1}{\sqrt{3}}$&&$l>0,\tilde qQ\geqslant0$&\eqref{eunint} at $\tilde q=0$\\
\cline{4-5}
$54.7^\circ<\left|\gamma_\mrm{KN}\ri|<60^\circ$&$\frac{1}{\sqrt{3}}>\frac{a}{M}>\frac{1}{2}$&&\multirow{3}{*}{$l>0,\tilde qQ>0$}\\
\cline{1-3}
$60^\circ$&$\frac{1}{2}$&{\bf 1c2a}&\\
\cline{1-3}
$60^\circ<\left|\gamma_\mrm{KN}\ri|<90^\circ$&$\frac{1}{2}>\frac{a}{M}>0$&{\bf 1b2a}&\\
\cline{1-5}
$90^\circ$&$0$&{\bf 1b2a3}&$l=0,\tilde qQ>0$&\eqref{eunint},\eqref{vtxlq},\eqref{veffconst} at $l=0$
\end{tabular}
\lbl{tab:sum}
\end{table}

As a last point, we discussed unusual behaviour in the $Q\to0$ limit, when one can maintain the magnitude of electrostatic force by considering very large $\left|\tilde q\ri|$. We found that for very small $Q$, critical particles with $\varepsilon_\mrm{cr}\leqslant1$ can have enormous values not only of specific charge, but also of angular momentum, and still be able to approach $r=M$. This is not possible for the cases $Q=0$ or $Q\approx a$. We discussed that this mega-BSW effect could have some significance in astrophysics because black holes can maintain a small charge due to interaction with external fields (see \cite{Wald74, Hiscock81}). However, as no black holes are expected to be precisely extremal, one should look for some analogues of this mega-BSW behaviour for nearly extremal black holes. This is left for future work.

\section*{Acknowledgements}

F.H. would like to thank Professor O. B. Zaslavskii for interesting discussions and for many inspiring suggestions that helped this work to evolve to its current stage. He also appreciates the goodwill of his colleagues from IDPASC and CENTRA, in particular its president J. P. S. Lemos. 
The work was supported by Fundação para a Ciência e a Tecnologia (Portugal) Grant No. PD/BD/113477/2015 awarded in the framework of the Doctoral Programme IDPASC-Portugal. F.H. is grateful for (especially travel) support from the Charles University, Grants No. GAUK 196516 and No. GAUK 606412. J.B. acknowledges support from Grant No. GAČR 14-37086G of the Czech Science Foundation (Albert Einstein Centre).

\appendix

\section{Auxiliary calculations}

\subsection{Derivatives of effective potentials}

\lbl{app:derwveff}

The turning point, which is also a stationary point of an effective potential, corresponds to a circular orbit. The effective potential in question may be either $W$ \cite{Wald, BarPrTeu} or $V$ \cite{Bardeen72} (see equations \eqref{weffpos} to \eqref{veffgen}). Let us examine the correspondence between the conditions, which holds for $r>r_+$ (or $N^2>0$, more precisely). Taking the radial derivative of \eqref{wveff}, we see
\rov{\pder{W}{r}=-\pder{V_+}{r}\(\varepsilon-V_-\)-\(\varepsilon-V_+\)\pder{V_-}{r}\rvt}{wveffder}
Indeed, all radial turning points indicated by $W$, which are also radial stationary points of $W$, are radial stationary points of either $V_+$ or $V_-$ as well,
\rov{W=0\conj\pder{W}{r}=0\equival\(\varepsilon=V_+\conj\pder{V_+}{r}=0\)\textrm{or}\(\varepsilon=V_-\conj\pder{V_-}{r}=0\)\rvt}{}
Thus, under restriction to the motion forward in time \eqref{fwpos}, $W$ and $V\equiv V_+$ are interchangeable for finding orbits.

The circular orbit, which is also an inflection point of an effective potential, is marginally stable. To see that $W$ and $V$ are interchangeable in this regard as well (cf. \cite{Wald, BarPrTeu, Bardeen72}), let us take another radial derivative of \eqref{wveff},
\rov{\pder[2]{W}{r}=-\pder[2]{V_+}{r}\(\varepsilon-V_-\)+2\pder{V_+}{r}\pder{V_-}{r}-\(\varepsilon-V_+\)\pder[2]{V_-}{r}\rvc}{wveffdder}
which leads to the desired conclusion
\rov{W=0\conj\pder{W}{r}=0\conj\pder[2]{W}{r}=0\equival\(\varepsilon=V_+\conj\pder{V_+}{r}=0\conj\pder[2]{V_+}{r}=0\)\textrm{or}\(\varepsilon=V_-\conj\pder{V_-}{r}=0\conj\pder[2]{V_-}{r}=0\)\rvt}{}

However, the most important result is an insight on how these results break down for $r\to r_+$, where $V_+\to V_-$ and derivatives of $V_\pm$ generally may not be finite (so $W$ may seem favourable). Nonetheless, for critical particles in extremal black hole spacetimes, a different form of correspondence emerges, and since (the radial derivative of) $V$ still embodies information about motion forward in time, it becomes preferable.

\subsection{The decomposition \eqref{regsing}}
\lbl{app:regsing}

Now let us present the general form of the contributions to \eqref{l2param} and \eqref{q2param} according to the decomposition \eqref{regsing}. Introducing the abbreviations
\prov{\msc W&=\dpder{\omega}{r}A_\varphi+\dpder{\phi}{r}\rvc&\msc N&=2\pder{\tilde N}{r}-\frac{\tilde N}{g_{\varphi\varphi}}\pder{g_{\varphi\varphi}}{r}\rvc}
the finite and the singular part of \eqref{q2param} can be written as
\rov{\tilde q_\mrm{reg}=-\res{\frac{\(\msc W\dpder{\omega}{r}+2\msc N\frac{\tilde N}{g_{\varphi\varphi}}\pder{A_\varphi}{r}\)\lambda A_\varphi-\(\msc{WN}+2\tilde N\dpder{\omega}{r}\pder{A_\varphi}{r}\)\sqrt{1+\frac{\lambda^2A_\varphi^2}{g_{\varphi\varphi}}}}{\frac{4{\tilde N}^2}{g_{\varphi\varphi}}\(\pder{A_\varphi}{r}\)^2-\msc W^2}}{r=r_0,\vartheta=\frac{\rpi}{2}}\rvc}{q2reg}
and
\rov{\tilde q_\mrm{sing}=-\res{\frac{2\msc W\tilde N\dpder{\omega}{r}\pder{A_\varphi}{r}-\msc W^2\frac{\tilde N}{g_{\varphi\varphi}}\pder{g_{\varphi\varphi}}{r}+\frac{8{\tilde N}^2}{g_{\varphi\varphi}}\pder{\tilde N}{r}\(\pder{A_\varphi}{r}\)^2}{\[\frac{4{\tilde N}^2}{g_{\varphi\varphi}}\(\pder{A_\varphi}{r}\)^2-\msc W^2\]\(\msc W\sqrt{1+\frac{\lambda^2A_\varphi^2}{g_{\varphi\varphi}}}+2\tilde N\frac{\lambda A_\varphi}{g_{\varphi\varphi}}\pder{A_\varphi}{r}\)}}{r=r_0,\vartheta=\frac{\rpi}{2}}\rvt}{q2sing}
Then, the expressions for contributions to \eqref{l2param} are closely related to the above,
\prov{l_\mrm{reg}&=\(\tilde q_\mrm{reg}-\lambda\)\res{A_\varphi}{r=r_0,\vartheta=\frac{\rpi}{2}}\rvc&l_\mrm{sing}&=\tilde q_\mrm{sing}\res{A_\varphi}{r=r_0,\vartheta=\frac{\rpi}{2}}\rvt\lbl{l2regsing}}

\end{document}

%% file: KN60DVeff.tex
\begingroup
  \makeatletter
  \providecommand\color[2][]{%
    \GenericError{(gnuplot) \space\space\space\@spaces}{%
      Package color not loaded in conjunction with
      terminal option `colourtext'%
    }{See the gnuplot documentation for explanation.%
    }{Either use 'blacktext' in gnuplot or load the package
      color.sty in LaTeX.}%
    \renewcommand\color[2][]{}%
  }%
  \providecommand\includegraphics[2][]{%
    \GenericError{(gnuplot) \space\space\space\@spaces}{%
      Package graphicx or graphics not loaded%
    }{See the gnuplot documentation for explanation.%
    }{The gnuplot epslatex terminal needs graphicx.sty or graphics.sty.}%
    \renewcommand\includegraphics[2][]{}%
  }%
  \providecommand\rotatebox[2]{#2}%
  \@ifundefined{ifGPcolor}{%
    \newif\ifGPcolor
    \GPcolortrue
  }{}%
  \@ifundefined{ifGPblacktext}{%
    \newif\ifGPblacktext
    \GPblacktexttrue
  }{}%
  \let\gplgaddtomacro\g@addto@macro
  \gdef\gplbacktext{}%
  \gdef\gplfronttext{}%
  \makeatother
  \ifGPblacktext
    \def\colorrgb#1{}%
    \def\colorgray#1{}%
  \else
    \ifGPcolor
      \def\colorrgb#1{\color[rgb]{#1}}%
      \def\colorgray#1{\color[gray]{#1}}%
      \expandafter\def\csname LTw\endcsname{\color{white}}%
      \expandafter\def\csname LTb\endcsname{\color{black}}%
      \expandafter\def\csname LTa\endcsname{\color{black}}%
      \expandafter\def\csname LT0\endcsname{\color[rgb]{1,0,0}}%
      \expandafter\def\csname LT1\endcsname{\color[rgb]{0,1,0}}%
      \expandafter\def\csname LT2\endcsname{\color[rgb]{0,0,1}}%
      \expandafter\def\csname LT3\endcsname{\color[rgb]{1,0,1}}%
      \expandafter\def\csname LT4\endcsname{\color[rgb]{0,1,1}}%
      \expandafter\def\csname LT5\endcsname{\color[rgb]{1,1,0}}%
      \expandafter\def\csname LT6\endcsname{\color[rgb]{0,0,0}}%
      \expandafter\def\csname LT7\endcsname{\color[rgb]{1,0.3,0}}%
      \expandafter\def\csname LT8\endcsname{\color[rgb]{0.5,0.5,0.5}}%
    \else
      \def\colorrgb#1{\color{black}}%
      \def\colorgray#1{\color[gray]{#1}}%
      \expandafter\def\csname LTw\endcsname{\color{white}}%
      \expandafter\def\csname LTb\endcsname{\color{black}}%
      \expandafter\def\csname LTa\endcsname{\color{black}}%
      \expandafter\def\csname LT0\endcsname{\color{black}}%
      \expandafter\def\csname LT1\endcsname{\color{black}}%
      \expandafter\def\csname LT2\endcsname{\color{black}}%
      \expandafter\def\csname LT3\endcsname{\color{black}}%
      \expandafter\def\csname LT4\endcsname{\color{black}}%
      \expandafter\def\csname LT5\endcsname{\color{black}}%
      \expandafter\def\csname LT6\endcsname{\color{black}}%
      \expandafter\def\csname LT7\endcsname{\color{black}}%
      \expandafter\def\csname LT8\endcsname{\color{black}}%
    \fi
  \fi
  \setlength{\unitlength}{0.0500bp}%
  \begin{picture}(7936.00,5668.00)%
    \gplgaddtomacro\gplbacktext{%
      \csname LTb\endcsname%
      \put(1078,704){\makebox(0,0)[r]{\strut{} 0.85}}%
      \put(1078,1644){\makebox(0,0)[r]{\strut{} 0.9}}%
      \put(1078,2584){\makebox(0,0)[r]{\strut{} 0.95}}%
      \put(1078,3523){\makebox(0,0)[r]{\strut{} 1}}%
      \put(1078,4463){\makebox(0,0)[r]{\strut{} 1.05}}%
      \put(1078,5403){\makebox(0,0)[r]{\strut{} 1.1}}%
      \put(1210,484){\makebox(0,0){\strut{} 1}}%
      \put(2001,484){\makebox(0,0){\strut{} 1.5}}%
      \put(2792,484){\makebox(0,0){\strut{} 2}}%
      \put(3583,484){\makebox(0,0){\strut{} 2.5}}%
      \put(4375,484){\makebox(0,0){\strut{} 3}}%
      \put(5166,484){\makebox(0,0){\strut{} 3.5}}%
      \put(5957,484){\makebox(0,0){\strut{} 4}}%
      \put(6748,484){\makebox(0,0){\strut{} 4.5}}%
      \put(7539,484){\makebox(0,0){\strut{} 5}}%
      \put(176,3053){\rotatebox{-270}{\makebox(0,0){\strut{}$V$}}}%
      \put(4374,154){\makebox(0,0){\strut{}$\frac{r}{M}$}}%
    }%
    \gplgaddtomacro\gplfronttext{%
      \csname LTb\endcsname%
      \put(4641,5230){\makebox(0,0)[r]{\strut{}$\tilde q\doteq0.33$}}%
      \csname LTb\endcsname%
      \put(4641,5010){\makebox(0,0)[r]{\strut{}$\tilde q\doteq0.39$}}%
      \csname LTb\endcsname%
      \put(4641,4790){\makebox(0,0)[r]{\strut{}$\tilde q=0.5$}}%
      \csname LTb\endcsname%
      \put(4641,4570){\makebox(0,0)[r]{\strut{}$\tilde q\doteq0.67$}}%
      \csname LTb\endcsname%
      \put(6552,5230){\makebox(0,0)[r]{\strut{}$\tilde q=0.75$}}%
      \csname LTb\endcsname%
      \put(6552,5010){\makebox(0,0)[r]{\strut{}$\tilde q=1$}}%
      \csname LTb\endcsname%
      \put(6552,4790){\makebox(0,0)[r]{\strut{}$\tilde q\doteq1.15$}}%
      \csname LTb\endcsname%
      \put(6552,4570){\makebox(0,0)[r]{\strut{}$\tilde q=1.25$}}%
    }%
    \gplbacktext
    \put(0,0){\includegraphics{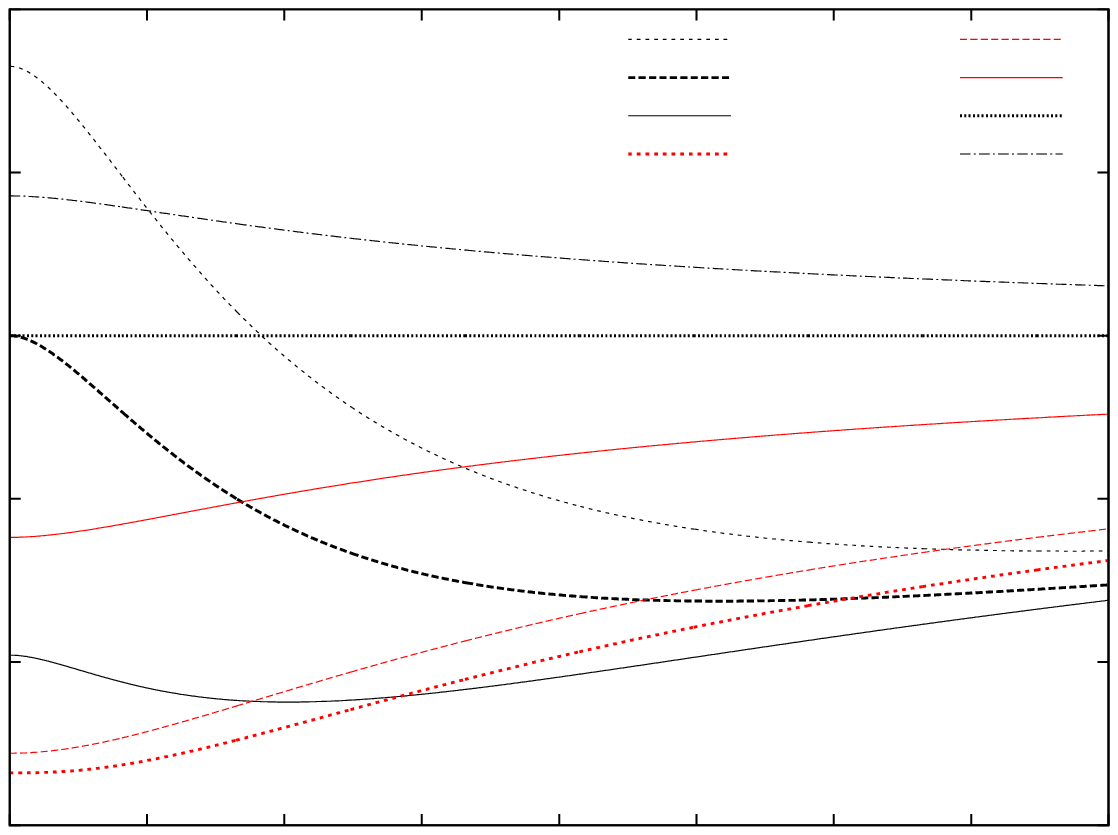}}%
    \gplfronttext
  \end{picture}%
\endgroup

%% file: exKNeqPS_1_2.tex
\begingroup
  \makeatletter
  \providecommand\color[2][]{%
    \GenericError{(gnuplot) \space\space\space\@spaces}{%
      Package color not loaded in conjunction with
      terminal option `colourtext'%
    }{See the gnuplot documentation for explanation.%
    }{Either use 'blacktext' in gnuplot or load the package
      color.sty in LaTeX.}%
    \renewcommand\color[2][]{}%
  }%
  \providecommand\includegraphics[2][]{%
    \GenericError{(gnuplot) \space\space\space\@spaces}{%
      Package graphicx or graphics not loaded%
    }{See the gnuplot documentation for explanation.%
    }{The gnuplot epslatex terminal needs graphicx.sty or graphics.sty.}%
    \renewcommand\includegraphics[2][]{}%
  }%
  \providecommand\rotatebox[2]{#2}%
  \@ifundefined{ifGPcolor}{%
    \newif\ifGPcolor
    \GPcolortrue
  }{}%
  \@ifundefined{ifGPblacktext}{%
    \newif\ifGPblacktext
    \GPblacktexttrue
  }{}%
  \let\gplgaddtomacro\g@addto@macro
  \gdef\gplbacktext{}%
  \gdef\gplfronttext{}%
  \makeatother
  \ifGPblacktext
    \def\colorrgb#1{}%
    \def\colorgray#1{}%
  \else
    \ifGPcolor
      \def\colorrgb#1{\color[rgb]{#1}}%
      \def\colorgray#1{\color[gray]{#1}}%
      \expandafter\def\csname LTw\endcsname{\color{white}}%
      \expandafter\def\csname LTb\endcsname{\color{black}}%
      \expandafter\def\csname LTa\endcsname{\color{black}}%
      \expandafter\def\csname LT0\endcsname{\color[rgb]{1,0,0}}%
      \expandafter\def\csname LT1\endcsname{\color[rgb]{0,1,0}}%
      \expandafter\def\csname LT2\endcsname{\color[rgb]{0,0,1}}%
      \expandafter\def\csname LT3\endcsname{\color[rgb]{1,0,1}}%
      \expandafter\def\csname LT4\endcsname{\color[rgb]{0,1,1}}%
      \expandafter\def\csname LT5\endcsname{\color[rgb]{1,1,0}}%
      \expandafter\def\csname LT6\endcsname{\color[rgb]{0,0,0}}%
      \expandafter\def\csname LT7\endcsname{\color[rgb]{1,0.3,0}}%
      \expandafter\def\csname LT8\endcsname{\color[rgb]{0.5,0.5,0.5}}%
    \else
      \def\colorrgb#1{\color{black}}%
      \def\colorgray#1{\color[gray]{#1}}%
      \expandafter\def\csname LTw\endcsname{\color{white}}%
      \expandafter\def\csname LTb\endcsname{\color{black}}%
      \expandafter\def\csname LTa\endcsname{\color{black}}%
      \expandafter\def\csname LT0\endcsname{\color{black}}%
      \expandafter\def\csname LT1\endcsname{\color{black}}%
      \expandafter\def\csname LT2\endcsname{\color{black}}%
      \expandafter\def\csname LT3\endcsname{\color{black}}%
      \expandafter\def\csname LT4\endcsname{\color{black}}%
      \expandafter\def\csname LT5\endcsname{\color{black}}%
      \expandafter\def\csname LT6\endcsname{\color{black}}%
      \expandafter\def\csname LT7\endcsname{\color{black}}%
      \expandafter\def\csname LT8\endcsname{\color{black}}%
    \fi
  \fi
  \setlength{\unitlength}{0.0500bp}%
  \begin{picture}(7936.00,5668.00)%
    \gplgaddtomacro\gplbacktext{%
      \csname LTb\endcsname%
      \put(682,704){\makebox(0,0)[r]{\strut{}$-2$}}%
      \put(682,1226){\makebox(0,0)[r]{\strut{}$-1$}}%
      \put(682,1748){\makebox(0,0)[r]{\strut{}$0$}}%
      \put(682,2270){\makebox(0,0)[r]{\strut{}$1$}}%
      \put(682,2792){\makebox(0,0)[r]{\strut{}$2$}}%
      \put(682,3315){\makebox(0,0)[r]{\strut{}$3$}}%
      \put(682,3837){\makebox(0,0)[r]{\strut{}$4$}}%
      \put(682,4359){\makebox(0,0)[r]{\strut{}$5$}}%
      \put(682,4881){\makebox(0,0)[r]{\strut{}$6$}}%
      \put(682,5403){\makebox(0,0)[r]{\strut{}$7$}}%
      \put(1168,484){\makebox(0,0){\strut{}$0$}}%
      \put(1876,484){\makebox(0,0){\strut{}$1$}}%
      \put(2584,484){\makebox(0,0){\strut{}$2$}}%
      \put(3292,484){\makebox(0,0){\strut{}$3$}}%
      \put(4000,484){\makebox(0,0){\strut{}$4$}}%
      \put(4707,484){\makebox(0,0){\strut{}$5$}}%
      \put(5415,484){\makebox(0,0){\strut{}$6$}}%
      \put(6123,484){\makebox(0,0){\strut{}$7$}}%
      \put(6831,484){\makebox(0,0){\strut{}$8$}}%
      \put(7539,484){\makebox(0,0){\strut{}$9$}}%
      \csname LTb\endcsname%
      \put(176,3053){\rotatebox{-270}{\makebox(0,0){\strut{}$\frac{l}{M}$}}}%
      \put(4176,154){\makebox(0,0){\strut{}$\tilde q$}}%
      \put(1770,2584){\rotatebox{-52}{\makebox(0,0)[l]{\strut{}$\varepsilon_\mrm{cr}=1$}}}%
      \put(2478,3001){\rotatebox{-52}{\makebox(0,0)[l]{\strut{}$\varepsilon_\mrm{cr}=2$}}}%
      \put(3185,3419){\rotatebox{-52}{\makebox(0,0)[l]{\strut{}$\varepsilon_\mrm{cr}=3$}}}%
    }%
    \gplgaddtomacro\gplfronttext{%
      \csname LTb\endcsname%
      \put(6552,5203){\makebox(0,0)[r]{\strut{}border ($\res{\nicepder{V}{r}}{r_0}=0$)}}%
      \csname LTb\endcsname%
      \put(6552,4928){\makebox(0,0)[r]{\strut{}asymptotes \eqref{admhypaskn}}}%
      \csname LTb\endcsname%
      \put(6552,4653){\makebox(0,0)[r]{\strut{}critical energy lines}}%
      \csname LTb\endcsname%
      \put(6552,4378){\makebox(0,0)[r]{\strut{}$\res{\nicedpder{V}{r}}{r_0}=0$}}%
      \csname LTb\endcsname%
      \put(6552,4103){\makebox(0,0)[r]{\strut{}asymptotes \eqref{dderas}}}%
    }%
    \gplbacktext
    \put(0,0){\includegraphics{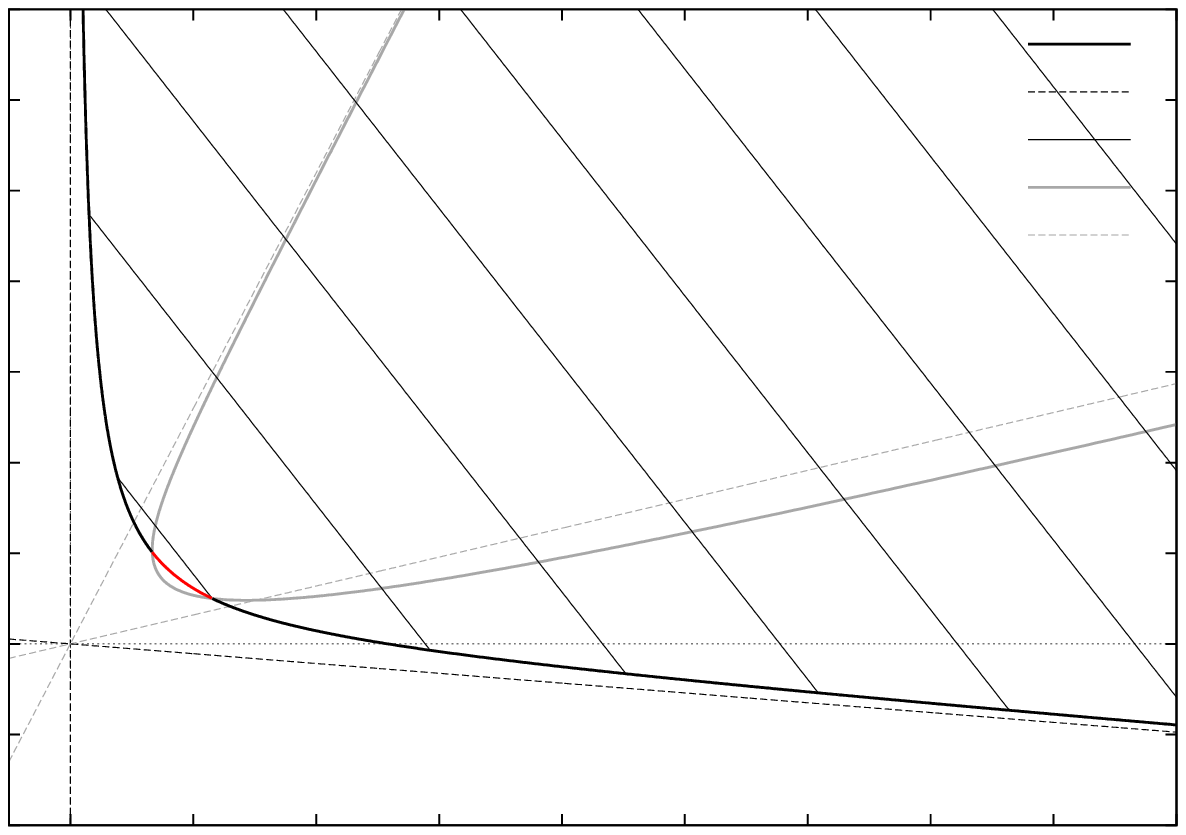}}%
    \gplfronttext
  \end{picture}%
\endgroup

%% file: exKNeqPS_s5-1_2.tex
\begingroup
  \makeatletter
  \providecommand\color[2][]{%
    \GenericError{(gnuplot) \space\space\space\@spaces}{%
      Package color not loaded in conjunction with
      terminal option `colourtext'%
    }{See the gnuplot documentation for explanation.%
    }{Either use 'blacktext' in gnuplot or load the package
      color.sty in LaTeX.}%
    \renewcommand\color[2][]{}%
  }%
  \providecommand\includegraphics[2][]{%
    \GenericError{(gnuplot) \space\space\space\@spaces}{%
      Package graphicx or graphics not loaded%
    }{See the gnuplot documentation for explanation.%
    }{The gnuplot epslatex terminal needs graphicx.sty or graphics.sty.}%
    \renewcommand\includegraphics[2][]{}%
  }%
  \providecommand\rotatebox[2]{#2}%
  \@ifundefined{ifGPcolor}{%
    \newif\ifGPcolor
    \GPcolortrue
  }{}%
  \@ifundefined{ifGPblacktext}{%
    \newif\ifGPblacktext
    \GPblacktexttrue
  }{}%
  \let\gplgaddtomacro\g@addto@macro
  \gdef\gplbacktext{}%
  \gdef\gplfronttext{}%
  \makeatother
  \ifGPblacktext
    \def\colorrgb#1{}%
    \def\colorgray#1{}%
  \else
    \ifGPcolor
      \def\colorrgb#1{\color[rgb]{#1}}%
      \def\colorgray#1{\color[gray]{#1}}%
      \expandafter\def\csname LTw\endcsname{\color{white}}%
      \expandafter\def\csname LTb\endcsname{\color{black}}%
      \expandafter\def\csname LTa\endcsname{\color{black}}%
      \expandafter\def\csname LT0\endcsname{\color[rgb]{1,0,0}}%
      \expandafter\def\csname LT1\endcsname{\color[rgb]{0,1,0}}%
      \expandafter\def\csname LT2\endcsname{\color[rgb]{0,0,1}}%
      \expandafter\def\csname LT3\endcsname{\color[rgb]{1,0,1}}%
      \expandafter\def\csname LT4\endcsname{\color[rgb]{0,1,1}}%
      \expandafter\def\csname LT5\endcsname{\color[rgb]{1,1,0}}%
      \expandafter\def\csname LT6\endcsname{\color[rgb]{0,0,0}}%
      \expandafter\def\csname LT7\endcsname{\color[rgb]{1,0.3,0}}%
      \expandafter\def\csname LT8\endcsname{\color[rgb]{0.5,0.5,0.5}}%
    \else
      \def\colorrgb#1{\color{black}}%
      \def\colorgray#1{\color[gray]{#1}}%
      \expandafter\def\csname LTw\endcsname{\color{white}}%
      \expandafter\def\csname LTb\endcsname{\color{black}}%
      \expandafter\def\csname LTa\endcsname{\color{black}}%
      \expandafter\def\csname LT0\endcsname{\color{black}}%
      \expandafter\def\csname LT1\endcsname{\color{black}}%
      \expandafter\def\csname LT2\endcsname{\color{black}}%
      \expandafter\def\csname LT3\endcsname{\color{black}}%
      \expandafter\def\csname LT4\endcsname{\color{black}}%
      \expandafter\def\csname LT5\endcsname{\color{black}}%
      \expandafter\def\csname LT6\endcsname{\color{black}}%
      \expandafter\def\csname LT7\endcsname{\color{black}}%
      \expandafter\def\csname LT8\endcsname{\color{black}}%
    \fi
  \fi
  \setlength{\unitlength}{0.0500bp}%
  \begin{picture}(7936.00,5668.00)%
    \gplgaddtomacro\gplbacktext{%
      \csname LTb\endcsname%
      \put(682,704){\makebox(0,0)[r]{\strut{}$-1$}}%
      \put(682,1226){\makebox(0,0)[r]{\strut{}$0$}}%
      \put(682,1748){\makebox(0,0)[r]{\strut{}$1$}}%
      \put(682,2270){\makebox(0,0)[r]{\strut{}$2$}}%
      \put(682,2792){\makebox(0,0)[r]{\strut{}$3$}}%
      \put(682,3315){\makebox(0,0)[r]{\strut{}$4$}}%
      \put(682,3837){\makebox(0,0)[r]{\strut{}$5$}}%
      \put(682,4359){\makebox(0,0)[r]{\strut{}$6$}}%
      \put(682,4881){\makebox(0,0)[r]{\strut{}$7$}}%
      \put(682,5403){\makebox(0,0)[r]{\strut{}$8$}}%
      \put(1168,484){\makebox(0,0){\strut{}$-2$}}%
      \put(1876,484){\makebox(0,0){\strut{}$-1$}}%
      \put(2584,484){\makebox(0,0){\strut{}$0$}}%
      \put(3292,484){\makebox(0,0){\strut{}$1$}}%
      \put(4000,484){\makebox(0,0){\strut{}$2$}}%
      \put(4707,484){\makebox(0,0){\strut{}$3$}}%
      \put(5415,484){\makebox(0,0){\strut{}$4$}}%
      \put(6123,484){\makebox(0,0){\strut{}$5$}}%
      \put(6831,484){\makebox(0,0){\strut{}$6$}}%
      \put(7539,484){\makebox(0,0){\strut{}$7$}}%
      \csname LTb\endcsname%
      \put(176,3053){\rotatebox{-270}{\makebox(0,0){\strut{}$\frac{l}{M}$}}}%
      \put(4176,154){\makebox(0,0){\strut{}$\tilde q$}}%
      \put(3008,2270){\rotatebox{-42}{\makebox(0,0)[l]{\strut{}$\varepsilon_\mrm{cr}=1$}}}%
      \put(3575,2897){\rotatebox{-42}{\makebox(0,0)[l]{\strut{}$\varepsilon_\mrm{cr}=2$}}}%
      \put(4141,3523){\rotatebox{-42}{\makebox(0,0)[l]{\strut{}$\varepsilon_\mrm{cr}=3$}}}%
    }%
    \gplgaddtomacro\gplfronttext{%
      \csname LTb\endcsname%
      \put(6552,5203){\makebox(0,0)[r]{\strut{}border ($\res{\nicepder{V}{r}}{r_0}=0$)}}%
      \csname LTb\endcsname%
      \put(6552,4928){\makebox(0,0)[r]{\strut{}asymptotes \eqref{admhypaskn}}}%
      \csname LTb\endcsname%
      \put(6552,4653){\makebox(0,0)[r]{\strut{}critical energy lines}}%
      \csname LTb\endcsname%
      \put(6552,4378){\makebox(0,0)[r]{\strut{}$\res{\nicedpder{V}{r}}{r_0}=0$}}%
      \csname LTb\endcsname%
      \put(6552,4103){\makebox(0,0)[r]{\strut{}asymptotes \eqref{dderas}}}%
    }%
    \gplbacktext
    \put(0,0){\includegraphics{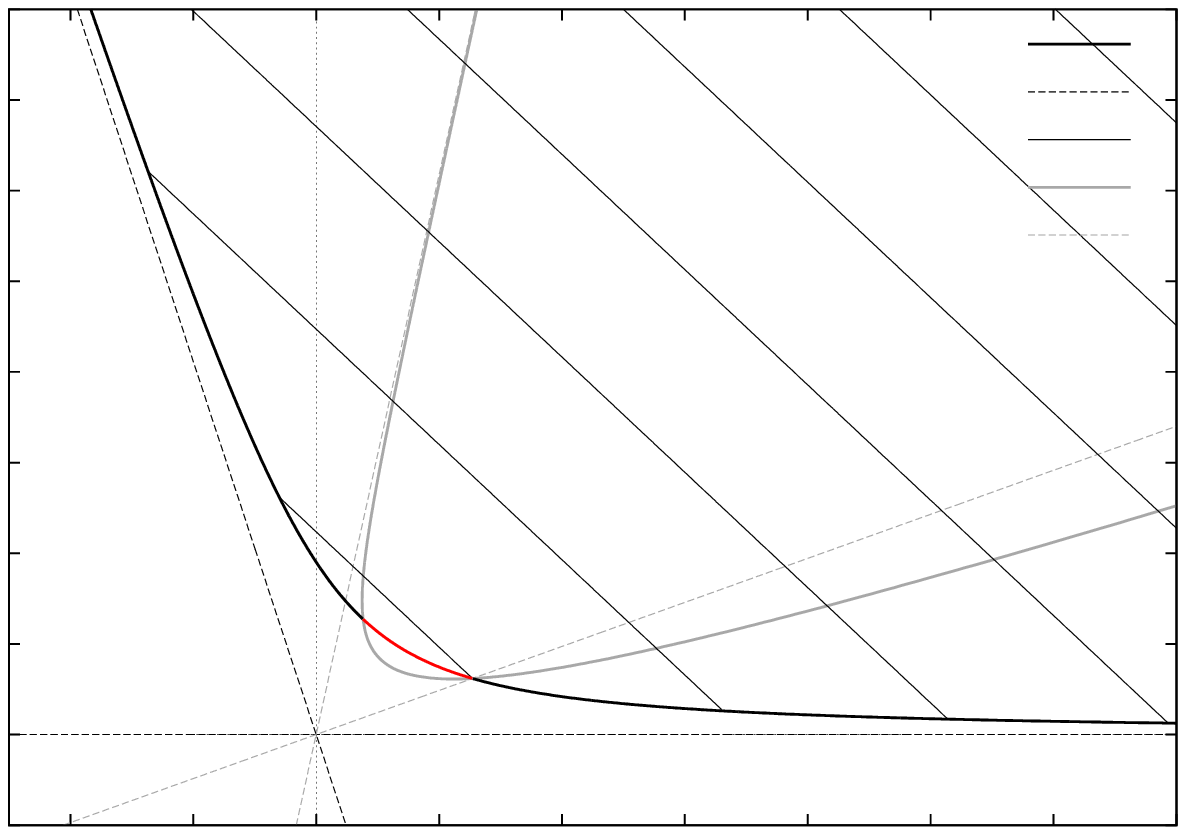}}%
    \gplfronttext
  \end{picture}%
\endgroup

%% file: exKNeqPS_0.tex
\begingroup
  \makeatletter
  \providecommand\color[2][]{%
    \GenericError{(gnuplot) \space\space\space\@spaces}{%
      Package color not loaded in conjunction with
      terminal option `colourtext'%
    }{See the gnuplot documentation for explanation.%
    }{Either use 'blacktext' in gnuplot or load the package
      color.sty in LaTeX.}%
    \renewcommand\color[2][]{}%
  }%
  \providecommand\includegraphics[2][]{%
    \GenericError{(gnuplot) \space\space\space\@spaces}{%
      Package graphicx or graphics not loaded%
    }{See the gnuplot documentation for explanation.%
    }{The gnuplot epslatex terminal needs graphicx.sty or graphics.sty.}%
    \renewcommand\includegraphics[2][]{}%
  }%
  \providecommand\rotatebox[2]{#2}%
  \@ifundefined{ifGPcolor}{%
    \newif\ifGPcolor
    \GPcolortrue
  }{}%
  \@ifundefined{ifGPblacktext}{%
    \newif\ifGPblacktext
    \GPblacktexttrue
  }{}%
  \let\gplgaddtomacro\g@addto@macro
  \gdef\gplbacktext{}%
  \gdef\gplfronttext{}%
  \makeatother
  \ifGPblacktext
    \def\colorrgb#1{}%
    \def\colorgray#1{}%
  \else
    \ifGPcolor
      \def\colorrgb#1{\color[rgb]{#1}}%
      \def\colorgray#1{\color[gray]{#1}}%
      \expandafter\def\csname LTw\endcsname{\color{white}}%
      \expandafter\def\csname LTb\endcsname{\color{black}}%
      \expandafter\def\csname LTa\endcsname{\color{black}}%
      \expandafter\def\csname LT0\endcsname{\color[rgb]{1,0,0}}%
      \expandafter\def\csname LT1\endcsname{\color[rgb]{0,1,0}}%
      \expandafter\def\csname LT2\endcsname{\color[rgb]{0,0,1}}%
      \expandafter\def\csname LT3\endcsname{\color[rgb]{1,0,1}}%
      \expandafter\def\csname LT4\endcsname{\color[rgb]{0,1,1}}%
      \expandafter\def\csname LT5\endcsname{\color[rgb]{1,1,0}}%
      \expandafter\def\csname LT6\endcsname{\color[rgb]{0,0,0}}%
      \expandafter\def\csname LT7\endcsname{\color[rgb]{1,0.3,0}}%
      \expandafter\def\csname LT8\endcsname{\color[rgb]{0.5,0.5,0.5}}%
    \else
      \def\colorrgb#1{\color{black}}%
      \def\colorgray#1{\color[gray]{#1}}%
      \expandafter\def\csname LTw\endcsname{\color{white}}%
      \expandafter\def\csname LTb\endcsname{\color{black}}%
      \expandafter\def\csname LTa\endcsname{\color{black}}%
      \expandafter\def\csname LT0\endcsname{\color{black}}%
      \expandafter\def\csname LT1\endcsname{\color{black}}%
      \expandafter\def\csname LT2\endcsname{\color{black}}%
      \expandafter\def\csname LT3\endcsname{\color{black}}%
      \expandafter\def\csname LT4\endcsname{\color{black}}%
      \expandafter\def\csname LT5\endcsname{\color{black}}%
      \expandafter\def\csname LT6\endcsname{\color{black}}%
      \expandafter\def\csname LT7\endcsname{\color{black}}%
      \expandafter\def\csname LT8\endcsname{\color{black}}%
    \fi
  \fi
  \setlength{\unitlength}{0.0500bp}%
  \begin{picture}(7936.00,5668.00)%
    \gplgaddtomacro\gplbacktext{%
      \csname LTb\endcsname%
      \put(682,965){\makebox(0,0)[r]{\strut{}$-4$}}%
      \put(682,1487){\makebox(0,0)[r]{\strut{}$-3$}}%
      \put(682,2009){\makebox(0,0)[r]{\strut{}$-2$}}%
      \put(682,2531){\makebox(0,0)[r]{\strut{}$-1$}}%
      \put(682,3054){\makebox(0,0)[r]{\strut{}$0$}}%
      \put(682,3576){\makebox(0,0)[r]{\strut{}$1$}}%
      \put(682,4098){\makebox(0,0)[r]{\strut{}$2$}}%
      \put(682,4620){\makebox(0,0)[r]{\strut{}$3$}}%
      \put(682,5142){\makebox(0,0)[r]{\strut{}$4$}}%
      \put(1168,484){\makebox(0,0){\strut{}$0$}}%
      \put(1876,484){\makebox(0,0){\strut{}$1$}}%
      \put(2584,484){\makebox(0,0){\strut{}$2$}}%
      \put(3292,484){\makebox(0,0){\strut{}$3$}}%
      \put(4000,484){\makebox(0,0){\strut{}$4$}}%
      \put(4707,484){\makebox(0,0){\strut{}$5$}}%
      \put(5415,484){\makebox(0,0){\strut{}$6$}}%
      \put(6123,484){\makebox(0,0){\strut{}$7$}}%
      \put(6831,484){\makebox(0,0){\strut{}$8$}}%
      \put(7539,484){\makebox(0,0){\strut{}$9$}}%
      \csname LTb\endcsname%
      \put(176,3053){\rotatebox{-270}{\makebox(0,0){\strut{}$\frac{l}{M}$}}}%
      \put(4176,154){\makebox(0,0){\strut{}$\tilde q$}}%
      \put(2725,3315){\rotatebox{-90}{\makebox(0,0)[l]{\strut{}$\varepsilon_\mrm{cr}=2$}}}%
      \put(3433,3315){\rotatebox{-90}{\makebox(0,0)[l]{\strut{}$\varepsilon_\mrm{cr}=3$}}}%
      \put(4141,3315){\rotatebox{-90}{\makebox(0,0)[l]{\strut{}$\varepsilon_\mrm{cr}=4$}}}%
    }%
    \gplgaddtomacro\gplfronttext{%
      \csname LTb\endcsname%
      \put(6542,3700){\makebox(0,0)[r]{\strut{}border ($\res{\nicepder{V}{r}}{r_0}=0$)}}%
      \csname LTb\endcsname%
      \put(6542,3425){\makebox(0,0)[r]{\strut{}critical energy lines}}%
      \csname LTb\endcsname%
      \put(6542,3150){\makebox(0,0)[r]{\strut{}asymptotes \eqref{admhypaskn}}}%
      \csname LTb\endcsname%
      \put(6542,2875){\makebox(0,0)[r]{\strut{}$\res{\nicedpder{V}{r}}{r_0}=0$}}%
      \csname LTb\endcsname%
      \put(6542,2600){\makebox(0,0)[r]{\strut{}asymptotes \eqref{dderas}}}%
    }%
    \gplbacktext
    \put(0,0){\includegraphics{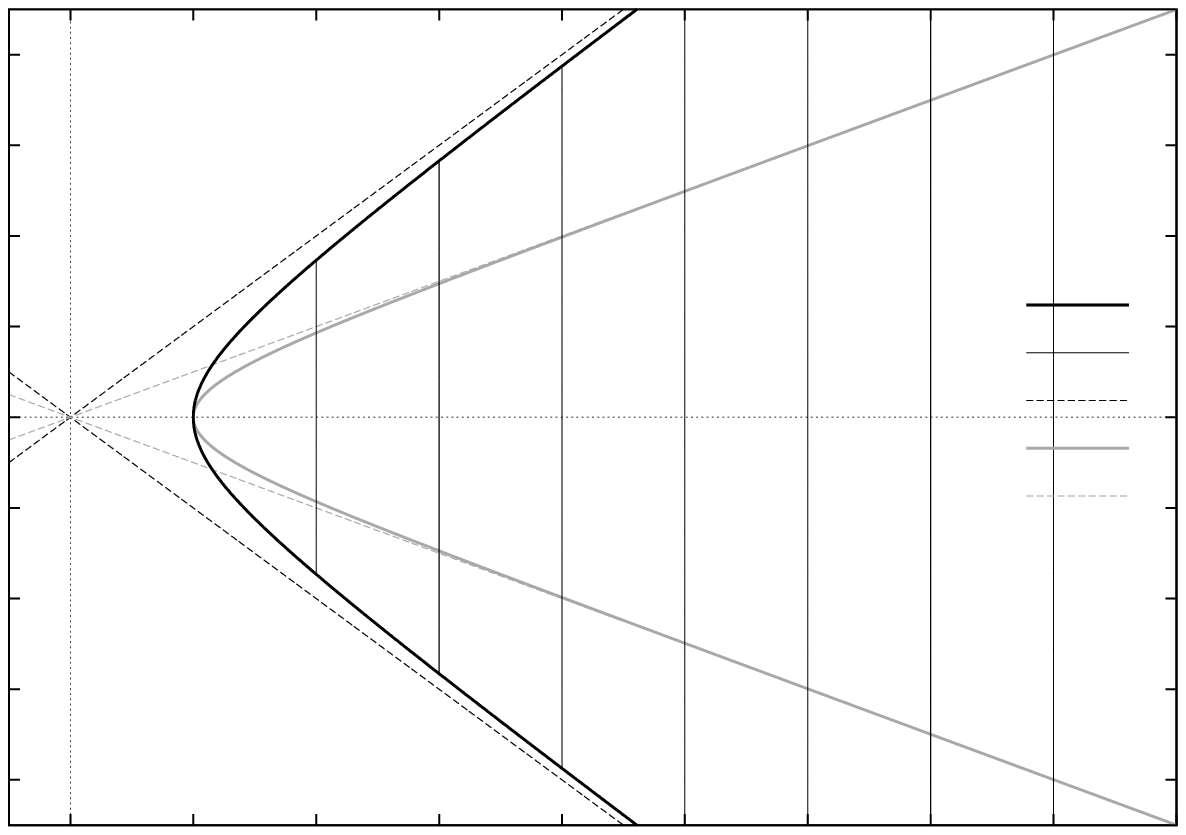}}%
    \gplfronttext
  \end{picture}%
\endgroup

%% file: exKNeqPS_1_s3.tex
\begingroup
  \makeatletter
  \providecommand\color[2][]{%
    \GenericError{(gnuplot) \space\space\space\@spaces}{%
      Package color not loaded in conjunction with
      terminal option `colourtext'%
    }{See the gnuplot documentation for explanation.%
    }{Either use 'blacktext' in gnuplot or load the package
      color.sty in LaTeX.}%
    \renewcommand\color[2][]{}%
  }%
  \providecommand\includegraphics[2][]{%
    \GenericError{(gnuplot) \space\space\space\@spaces}{%
      Package graphicx or graphics not loaded%
    }{See the gnuplot documentation for explanation.%
    }{The gnuplot epslatex terminal needs graphicx.sty or graphics.sty.}%
    \renewcommand\includegraphics[2][]{}%
  }%
  \providecommand\rotatebox[2]{#2}%
  \@ifundefined{ifGPcolor}{%
    \newif\ifGPcolor
    \GPcolortrue
  }{}%
  \@ifundefined{ifGPblacktext}{%
    \newif\ifGPblacktext
    \GPblacktexttrue
  }{}%
  \let\gplgaddtomacro\g@addto@macro
  \gdef\gplbacktext{}%
  \gdef\gplfronttext{}%
  \makeatother
  \ifGPblacktext
    \def\colorrgb#1{}%
    \def\colorgray#1{}%
  \else
    \ifGPcolor
      \def\colorrgb#1{\color[rgb]{#1}}%
      \def\colorgray#1{\color[gray]{#1}}%
      \expandafter\def\csname LTw\endcsname{\color{white}}%
      \expandafter\def\csname LTb\endcsname{\color{black}}%
      \expandafter\def\csname LTa\endcsname{\color{black}}%
      \expandafter\def\csname LT0\endcsname{\color[rgb]{1,0,0}}%
      \expandafter\def\csname LT1\endcsname{\color[rgb]{0,1,0}}%
      \expandafter\def\csname LT2\endcsname{\color[rgb]{0,0,1}}%
      \expandafter\def\csname LT3\endcsname{\color[rgb]{1,0,1}}%
      \expandafter\def\csname LT4\endcsname{\color[rgb]{0,1,1}}%
      \expandafter\def\csname LT5\endcsname{\color[rgb]{1,1,0}}%
      \expandafter\def\csname LT6\endcsname{\color[rgb]{0,0,0}}%
      \expandafter\def\csname LT7\endcsname{\color[rgb]{1,0.3,0}}%
      \expandafter\def\csname LT8\endcsname{\color[rgb]{0.5,0.5,0.5}}%
    \else
      \def\colorrgb#1{\color{black}}%
      \def\colorgray#1{\color[gray]{#1}}%
      \expandafter\def\csname LTw\endcsname{\color{white}}%
      \expandafter\def\csname LTb\endcsname{\color{black}}%
      \expandafter\def\csname LTa\endcsname{\color{black}}%
      \expandafter\def\csname LT0\endcsname{\color{black}}%
      \expandafter\def\csname LT1\endcsname{\color{black}}%
      \expandafter\def\csname LT2\endcsname{\color{black}}%
      \expandafter\def\csname LT3\endcsname{\color{black}}%
      \expandafter\def\csname LT4\endcsname{\color{black}}%
      \expandafter\def\csname LT5\endcsname{\color{black}}%
      \expandafter\def\csname LT6\endcsname{\color{black}}%
      \expandafter\def\csname LT7\endcsname{\color{black}}%
      \expandafter\def\csname LT8\endcsname{\color{black}}%
    \fi
  \fi
  \setlength{\unitlength}{0.0500bp}%
  \begin{picture}(7936.00,5668.00)%
    \gplgaddtomacro\gplbacktext{%
      \csname LTb\endcsname%
      \put(682,704){\makebox(0,0)[r]{\strut{}$-1$}}%
      \put(682,1226){\makebox(0,0)[r]{\strut{}$0$}}%
      \put(682,1748){\makebox(0,0)[r]{\strut{}$1$}}%
      \put(682,2270){\makebox(0,0)[r]{\strut{}$2$}}%
      \put(682,2792){\makebox(0,0)[r]{\strut{}$3$}}%
      \put(682,3315){\makebox(0,0)[r]{\strut{}$4$}}%
      \put(682,3837){\makebox(0,0)[r]{\strut{}$5$}}%
      \put(682,4359){\makebox(0,0)[r]{\strut{}$6$}}%
      \put(682,4881){\makebox(0,0)[r]{\strut{}$7$}}%
      \put(682,5403){\makebox(0,0)[r]{\strut{}$8$}}%
      \put(1168,484){\makebox(0,0){\strut{}$-1$}}%
      \put(1876,484){\makebox(0,0){\strut{}$0$}}%
      \put(2584,484){\makebox(0,0){\strut{}$1$}}%
      \put(3292,484){\makebox(0,0){\strut{}$2$}}%
      \put(4000,484){\makebox(0,0){\strut{}$3$}}%
      \put(4707,484){\makebox(0,0){\strut{}$4$}}%
      \put(5415,484){\makebox(0,0){\strut{}$5$}}%
      \put(6123,484){\makebox(0,0){\strut{}$6$}}%
      \put(6831,484){\makebox(0,0){\strut{}$7$}}%
      \put(7539,484){\makebox(0,0){\strut{}$8$}}%
      \csname LTb\endcsname%
      \put(176,3053){\rotatebox{-270}{\makebox(0,0){\strut{}$\frac{l}{M}$}}}%
      \put(4176,154){\makebox(0,0){\strut{}$\tilde q$}}%
      \put(2371,2218){\rotatebox{-47}{\makebox(0,0)[l]{\strut{}$\varepsilon_\mrm{cr}=1$}}}%
      \put(3079,2688){\rotatebox{-47}{\makebox(0,0)[l]{\strut{}$\varepsilon_\mrm{cr}=2$}}}%
      \put(3787,3106){\rotatebox{-47}{\makebox(0,0)[l]{\strut{}$\varepsilon_\mrm{cr}=3$}}}%
    }%
    \gplgaddtomacro\gplfronttext{%
      \csname LTb\endcsname%
      \put(6552,5203){\makebox(0,0)[r]{\strut{}border ($\res{\nicepder{V}{r}}{r_0}=0$)}}%
      \csname LTb\endcsname%
      \put(6552,4928){\makebox(0,0)[r]{\strut{}critical energy lines}}%
      \csname LTb\endcsname%
      \put(6552,4653){\makebox(0,0)[r]{\strut{}asymptotes \eqref{admhypaskn}}}%
      \csname LTb\endcsname%
      \put(6552,4378){\makebox(0,0)[r]{\strut{}$\res{\nicedpder{V}{r}}{r_0}=0$}}%
      \csname LTb\endcsname%
      \put(6552,4103){\makebox(0,0)[r]{\strut{}asymptotes \eqref{dderas}}}%
    }%
    \gplbacktext
    \put(0,0){\includegraphics{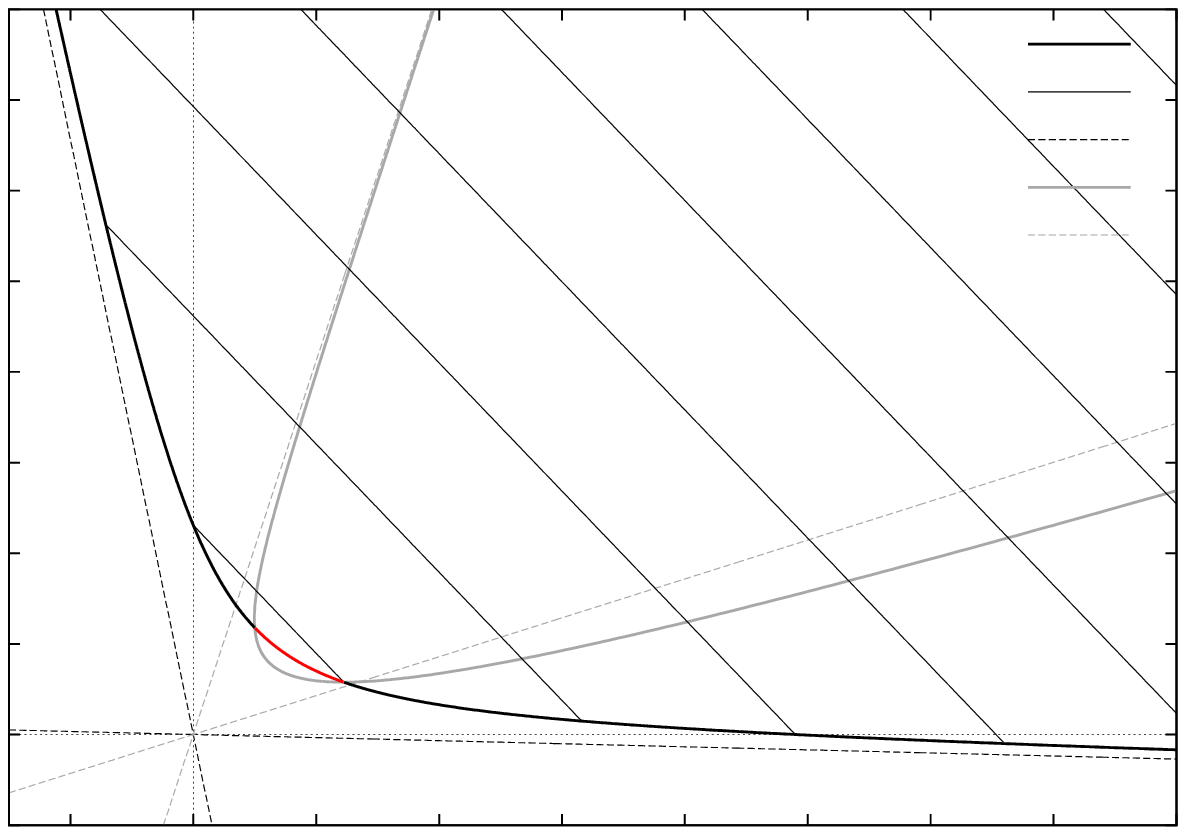}}%
    \gplfronttext
  \end{picture}%
\endgroup

%% file: exKNeqPS_1_s2.tex
\begingroup
  \makeatletter
  \providecommand\color[2][]{%
    \GenericError{(gnuplot) \space\space\space\@spaces}{%
      Package color not loaded in conjunction with
      terminal option `colourtext'%
    }{See the gnuplot documentation for explanation.%
    }{Either use 'blacktext' in gnuplot or load the package
      color.sty in LaTeX.}%
    \renewcommand\color[2][]{}%
  }%
  \providecommand\includegraphics[2][]{%
    \GenericError{(gnuplot) \space\space\space\@spaces}{%
      Package graphicx or graphics not loaded%
    }{See the gnuplot documentation for explanation.%
    }{The gnuplot epslatex terminal needs graphicx.sty or graphics.sty.}%
    \renewcommand\includegraphics[2][]{}%
  }%
  \providecommand\rotatebox[2]{#2}%
  \@ifundefined{ifGPcolor}{%
    \newif\ifGPcolor
    \GPcolortrue
  }{}%
  \@ifundefined{ifGPblacktext}{%
    \newif\ifGPblacktext
    \GPblacktexttrue
  }{}%
  \let\gplgaddtomacro\g@addto@macro
  \gdef\gplbacktext{}%
  \gdef\gplfronttext{}%
  \makeatother
  \ifGPblacktext
    \def\colorrgb#1{}%
    \def\colorgray#1{}%
  \else
    \ifGPcolor
      \def\colorrgb#1{\color[rgb]{#1}}%
      \def\colorgray#1{\color[gray]{#1}}%
      \expandafter\def\csname LTw\endcsname{\color{white}}%
      \expandafter\def\csname LTb\endcsname{\color{black}}%
      \expandafter\def\csname LTa\endcsname{\color{black}}%
      \expandafter\def\csname LT0\endcsname{\color[rgb]{1,0,0}}%
      \expandafter\def\csname LT1\endcsname{\color[rgb]{0,1,0}}%
      \expandafter\def\csname LT2\endcsname{\color[rgb]{0,0,1}}%
      \expandafter\def\csname LT3\endcsname{\color[rgb]{1,0,1}}%
      \expandafter\def\csname LT4\endcsname{\color[rgb]{0,1,1}}%
      \expandafter\def\csname LT5\endcsname{\color[rgb]{1,1,0}}%
      \expandafter\def\csname LT6\endcsname{\color[rgb]{0,0,0}}%
      \expandafter\def\csname LT7\endcsname{\color[rgb]{1,0.3,0}}%
      \expandafter\def\csname LT8\endcsname{\color[rgb]{0.5,0.5,0.5}}%
    \else
      \def\colorrgb#1{\color{black}}%
      \def\colorgray#1{\color[gray]{#1}}%
      \expandafter\def\csname LTw\endcsname{\color{white}}%
      \expandafter\def\csname LTb\endcsname{\color{black}}%
      \expandafter\def\csname LTa\endcsname{\color{black}}%
      \expandafter\def\csname LT0\endcsname{\color{black}}%
      \expandafter\def\csname LT1\endcsname{\color{black}}%
      \expandafter\def\csname LT2\endcsname{\color{black}}%
      \expandafter\def\csname LT3\endcsname{\color{black}}%
      \expandafter\def\csname LT4\endcsname{\color{black}}%
      \expandafter\def\csname LT5\endcsname{\color{black}}%
      \expandafter\def\csname LT6\endcsname{\color{black}}%
      \expandafter\def\csname LT7\endcsname{\color{black}}%
      \expandafter\def\csname LT8\endcsname{\color{black}}%
    \fi
  \fi
  \setlength{\unitlength}{0.0500bp}%
  \begin{picture}(7936.00,5668.00)%
    \gplgaddtomacro\gplbacktext{%
      \csname LTb\endcsname%
      \put(682,704){\makebox(0,0)[r]{\strut{}$-1$}}%
      \put(682,1226){\makebox(0,0)[r]{\strut{}$0$}}%
      \put(682,1748){\makebox(0,0)[r]{\strut{}$1$}}%
      \put(682,2270){\makebox(0,0)[r]{\strut{}$2$}}%
      \put(682,2792){\makebox(0,0)[r]{\strut{}$3$}}%
      \put(682,3315){\makebox(0,0)[r]{\strut{}$4$}}%
      \put(682,3837){\makebox(0,0)[r]{\strut{}$5$}}%
      \put(682,4359){\makebox(0,0)[r]{\strut{}$6$}}%
      \put(682,4881){\makebox(0,0)[r]{\strut{}$7$}}%
      \put(682,5403){\makebox(0,0)[r]{\strut{}$8$}}%
      \put(1487,484){\makebox(0,0){\strut{}$-2$}}%
      \put(2832,484){\makebox(0,0){\strut{}$0$}}%
      \put(4177,484){\makebox(0,0){\strut{}$2$}}%
      \put(5522,484){\makebox(0,0){\strut{}$4$}}%
      \put(6867,484){\makebox(0,0){\strut{}$6$}}%
      \csname LTb\endcsname%
      \put(176,3053){\rotatebox{-270}{\makebox(0,0){\strut{}$\frac{l}{M}$}}}%
      \put(4176,154){\makebox(0,0){\strut{}$\tilde q$}}%
      \put(3235,2270){\rotatebox{-38}{\makebox(0,0)[l]{\strut{}$\varepsilon_\mrm{cr}=1$}}}%
      \put(3773,2949){\rotatebox{-38}{\makebox(0,0)[l]{\strut{}$\varepsilon_\mrm{cr}=2$}}}%
      \put(4311,3628){\rotatebox{-38}{\makebox(0,0)[l]{\strut{}$\varepsilon_\mrm{cr}=3$}}}%
    }%
    \gplgaddtomacro\gplfronttext{%
      \csname LTb\endcsname%
      \put(6552,5203){\makebox(0,0)[r]{\strut{}border ($\res{\nicepder{V}{r}}{r_0}=0$)}}%
      \csname LTb\endcsname%
      \put(6552,4928){\makebox(0,0)[r]{\strut{}critical energy lines}}%
      \csname LTb\endcsname%
      \put(6552,4653){\makebox(0,0)[r]{\strut{}asymptotes \eqref{admhypaskn}}}%
      \csname LTb\endcsname%
      \put(6552,4378){\makebox(0,0)[r]{\strut{}$\res{\nicedpder{V}{r}}{r_0}=0$}}%
      \csname LTb\endcsname%
      \put(6552,4103){\makebox(0,0)[r]{\strut{}asymptotes \eqref{dderas}}}%
    }%
    \gplbacktext
    \put(0,0){\includegraphics{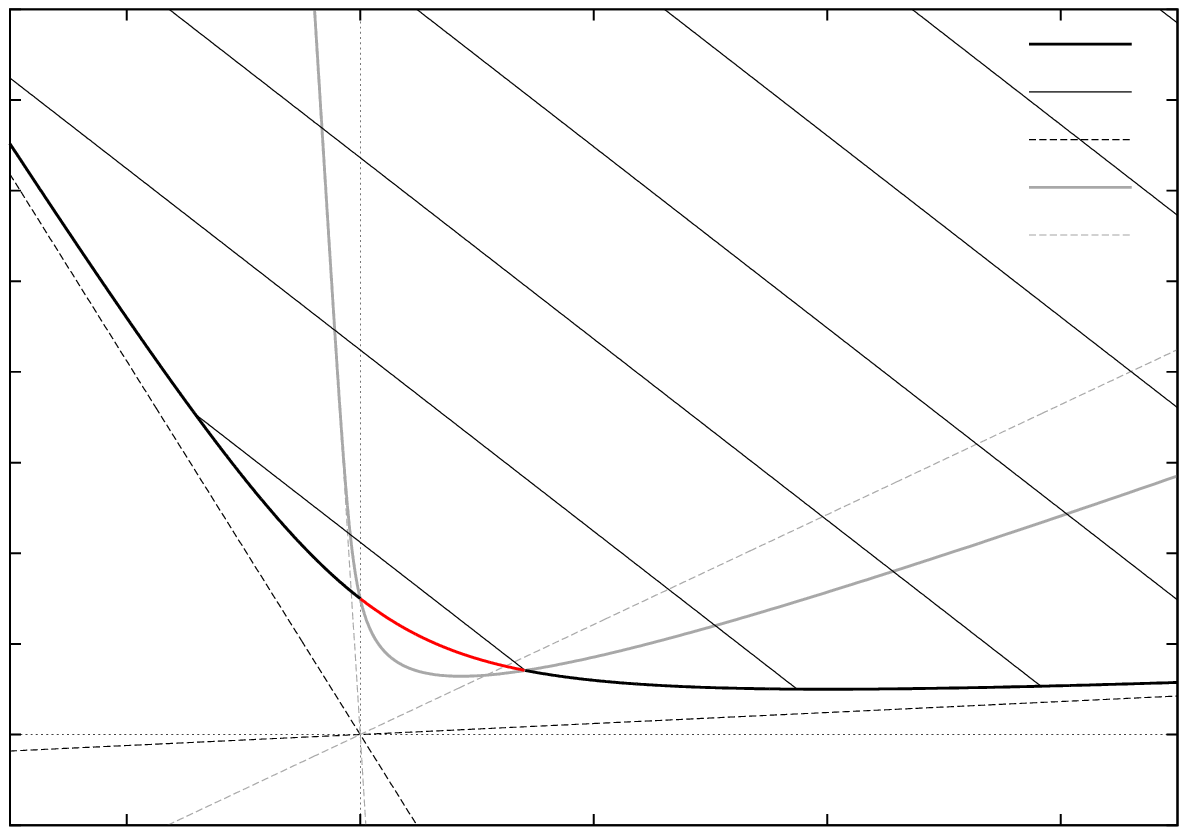}}%
    \gplfronttext
  \end{picture}%
\endgroup

%% file: Article-arXiv.bbl
\begin{thebibliography}{}

\bibitem{Penrose69}
  R. {Penrose}, 
  {Gravitational Collapse: the Role of General Relativity},
  Gen. Relat. Gravit. {\bf 34}, 1141-1165 (2002). Original edition: 
  Riv. Nuovo Cimento, Numero Speziale {\bf 1}, 252
  (1969).

\bibitem{Christd70}
  D. {Christodoulou},
  {Reversible and Irreversible Transformations in Black-Hole Physics},
  {Phys. Rev. Lett.} {\bf 25},
  {1596--1597} 
  (1970).

\bibitem{BarPrTeu}
  J. M. {Bardeen}, W. H. {Press}, S. A. {Teukolsky}, 
  {Rotating black holes: Locally nonrotating frames, energy extraction, and scalar synchrotron radiation}, 
  Astrophys. J.
  {\bf 178}, 347-370 
  (1972).

\bibitem{PirShK}
  T. {Piran}, J. {Shaham}, J. {Katz}, 
  {High efficiency of the Penrose mechanism for particle collisions},
  Astrophys. J. Lett. {\bf 196}, L107-L108 
  (1975).

\bibitem{PirSh}
  T. Piran, J. Shaham, 
  {Upper bounds on collisional Penrose processes near rotating black-hole horizons},
  {Phys. Rev. D} {\bf 16},
  {1615--1635} (1977).

\bibitem{BSW}
  M. {Bañados}, J. {Silk}, S. M. {West}, 
  {Kerr Black Holes as Particle Accelerators to Arbitrarily High Energy}, 
  Phys. Rev. Lett.
  {\bf 103},
  111102
  (2009), 4 pages,
  arXiv:0909.0169 [hep-ph].

\bibitem{Zasl11b}
  O. B. {Zaslavskii}, 
  {Acceleration of particles by black holes—a general explanation}.
  Classical Quant. Grav. {\bf 28}, 
  105010 (2011), 7 pages,
  arXiv:1011.0167 [gr-qc]. 

\bibitem{HK14}
  T. Harada, M. Kimura,
  {Black holes as particle accelerators: a brief review},
  {Classical and Quantum Gravity} {\bf 31},
  {243001} (2014), 17 pages,
  arXiv:1409.7502 [gr-qc].

\bibitem{Zasl12a}
  O. B. {Zaslavskii},
  {Acceleration of particles near the inner black hole horizon},
  {Phys. Rev. D} 
  {\bf 85},
  {024029}
  (2012),
  9 pages,
  arXiv:1110.5838 [gr-qc].

\bibitem{GP10a}
  A. A. {Grib}, Yu. V. {Pavlov}, 
  {On the collisions between particles in the vicinity of rotating black holes}, 
  JETP Letters  
  {\bf 92}, 125-129 
  (2010),
  arXiv:1004.0913 [gr-qc].

\bibitem{GP11a}
  A. A. Grib, Yu. V. Pavlov, On particle collisions near rotating black holes,
  Gravitation and Cosmology {\bf 17},
  42--46 (2011),
  arXiv:1010.2052 [gr-qc].

\bibitem{BCGPU}
  E. {Berti}, V. {Cardoso}, L. {Gualtieri}, F. {Pretorius}, U. {Sperhake},
  {Comment on ``Kerr Black Holes as Particle Accelerators to Arbitrarily High Energy''}, 
  Phys. Rev. Lett. {\bf 103}, 
  239001 (2009), 1 page,
  arXiv:0911.2243 [gr-qc].

\bibitem{JacSot}
  T. {Jacobson}, T. P. {Sotiriou}, 
  {Spinning Black Holes as Particle Accelerators}, 
  Phys. Rev. Lett.  {\bf 104},
  021101 (2010), 3 pages,
  arXiv:0911.3363 [gr-qc].

\bibitem{KimNakTag}
  M. Kimura, K.-i. Nakao, and H. Tagoshi,
{Acceleration of colliding shells around a black hole: Validity of the test particle approximation in the Banados-Silk-West process}, 
  Phys. Rev. D {\bf 83},
  044013 (2011), 8 pages,
  arXiv:1010.5438 [gr-qc]. 

\bibitem{TantZasl}
  I. V. {Tanatarov}, O. B. {Zaslavskii},
  {Ba\~nados-Silk-West effect with nongeodesic particles: Extremal horizons},
  {Phys. Rev. D} 
  {\bf 88},
  {064036}
  (2013),
  14 pages,
  arXiv:1307.0034 [gr-qc].  

\bibitem{HaNeMi}
  T. Harada, H. Nemoto, U. Miyamoto,
  {Upper limits of particle emission from high-energy collision and reaction near a maximally rotating Kerr black hole},
  {Phys. Rev. D} {\bf 86},
  {024027} (2012),
  10 pages,
  arXiv:1205.7088 [gr-qc].

\bibitem{Zasl12b}
  O. B. {Zaslavskii},
  {Energetics of particle collisions near dirty rotating extremal black holes: Banados-Silk-West effect versus Penrose process},
  {Phys. Rev. D} {\bf 86},
  {084030}
  (2012), 14 pages,
  arXiv:1205.4410 [gr-qc].

\bibitem{Zasl12c}
  O. B. {Zaslavskii}, {Energy extraction from extremal charged black holes due to the Banados-Silk-West effect},
  {Phys. Rev. D} {\bf 86},
  {124039} (2012), 4 pages,
  arXiv:1207.5209 [gr-qc]. 

\bibitem{BPAH}
  M. Bejger, T. Piran, M. Abramowicz, F. H\aa{}kanson,
  {Collisional Penrose Process near the Horizon of Extreme Kerr Black Holes},
  {Phys. Rev. Lett.} {\bf 109},
  {121101} 
  (2012), 5 pages,
  arXiv:1205.4350 [astro-ph.HE].

\bibitem{Schnitt14}
  J. D. Schnittman, 
  {Revised Upper Limit to Energy Extraction from a Kerr Black Hole},
  {Phys. Rev. Lett.} {\bf 113},
  {261102} (2014), 5 pages,
  arXiv:1410.6446 [astro-ph.HE].

\bibitem{BeBrCa}
  E. {Berti}, R. Brito, V. Cardoso,
  {Ultrahigh-Energy Debris from the Collisional Penrose Process},
  {Phys. Rev. Lett.} {\bf 114},
  {251103} (2015), 5 pages,
  arXiv:1410.8534 [gr-qc].
  
\bibitem{Zasl15b}
  O. B. Zaslavskii,
  {General limitations on trajectories suitable for super-Penrose process},
  Europhys. Lett. {\bf 111},
  {50004} 
  (2015), 4 pages,
  arXiv:1506.06527 [gr-qc].

\bibitem{BHSW}
  M. {Bañados}, B. {Hassanain}, J. {Silk}, S. M. {West}, 
  {Emergent flux from particle collisions near a Kerr black hole},
  Phys. Rev. D {\bf 83},
  023004 (2011), 10 pages,
  arXiv:1010.2724 [astro-ph.CO].

\bibitem{Schnitt15}
  J. D. Schnittman,
  {The Distribution and Annihilation of Dark Matter Around Black Holes},
  Astrophys. J. {\bf 806},
  {264} 
  (2015), 18 pages,
  arXiv:1506.06728 [astro-ph.HE].

\bibitem{HK11b}
  T. {Harada}, M. {Kimura},
  {Collision of two general geodesic particles around a Kerr black hole}, 
  Phys. Rev. D {\bf 83}
  084041 (2011), 9 pages,
  arXiv:1102.3316 [gr-qc].
  
\bibitem{Zasl12d}  
  O. B. Zaslavskii, 
  Ultra-high energy collisions of nonequatorial geodesic particles near dirty black holes,
  Journal of High Energy Physics 12 (2012) 32, 9 pages,
  arXiv:1209.4987 [gr-qc].

\bibitem{Liu13}
  Ch.-Q. Liu,
  {Collision of Two General Geodesic Particles around a Kerr—Newman Black Hole},
  {Chinese Physics Letters}
  {\bf 30},
  {100401}
  (2013), 5 pages.
    
\bibitem{Zasl10}
  O. B. {Zaslavskii}, 
  {Acceleration of particles as a universal property of rotating black holes}, 
  Phys. Rev. D 
  {\bf 82}, 
  083004 (2010), 5 pages,
  arXiv:1007.3678 [gr-qc].

\bibitem{WeiLiuGuoFu}
  S.-W. Wei, Y.-X. Liu, H. Guo, C.-E. Fu,
  {Charged spinning black holes as particle accelerators}, 
  Phys. Rev. D  {\bf 82},
  103005 (2010), 6 pages,
  arXiv:1006.1056 [hep-th].

\bibitem{Zasl11a}
  O. B. {Zaslavskii}, 
  {Acceleration of particles by nonrotating charged black holes?}
  JETP Letters 
  {\bf 92}, 571-574 
  (2010).

\bibitem{Wilkins}
  D. C. Wilkins,
  Bound Geodesics in the Kerr Metric,
  Phys. Rev. D {\bf 5}, 814-822
  (1972).

\bibitem{Bardeen72}
  J. M. {Bardeen}, 
  {Timelike and Null Geodesics in the Kerr Metric}, in
  \emph{Black holes (Les Houches lectures)},
  eds. B. DeWitt and C. DeWitt, 
  (Gordon and Breach, New York, 1972).

\bibitem{Article1}
  J. Bičák, F. Hejda,
  {Near-horizon description of extremal magnetized stationary
black holes and Meissner effect}, 
  {Phys. Rev. D} 
  {\bf 92}, 104006 (2015), 19 pages,
  arXiv:1510.01911 [gr-qc]. 

\bibitem{Frolov12}
  V. P. Frolov, 
  {Weakly magnetized black holes as particle accelerators},
  {Phys. Rev. D} {\bf 85},
  {024020} (2012), 6 pages,
  arXiv:1110.6274 [gr-qc].

\bibitem{IgHaKi}
  T. Igata, T. Harada, M. Kimura,
  {Effect of a weak electromagnetic field on particle acceleration by a rotating black hole},
  {Phys. Rev. D} {\bf 85},
  {104028}
  (2012), 11 pages,
  arXiv:1202.4859 [gr-qc].
  
\bibitem{Zasl15a} 
  O. B. Zaslavskii,
  Innermost stable circular orbit near dirty black holes in magnetic field and ultra-high-energy particle collisions,
  Eur. Phys. J. C {\bf 75}, 
  403 (2015), 14 pages,
  arXiv:1405.2543 [gr-qc].

\bibitem{dipl}
  F. {Hejda}, \emph{Particles and fields in curved spacetimes (selected problems)}, 
  Master thesis 
  (Institute of Theoretical Physics, Charles University in Prague, September 2013), 
  \url{https://is.cuni.cz/webapps/zzp/detail/117072/}.
  
\bibitem{FrolNov}
   V. P. {Frolov}, I. D. {Novikov}, 
  \emph{Black hole physics: basic concepts and new developments} 
  (Springer, 1998).

\bibitem{Sharp}
  N. A. Sharp, 
  {Geodesics in black hole space-times},
  {General Relativity and Gravitation} {\bf 10},
  659--670 (1979).

\bibitem{Wald}
  R. M. {Wald}, 
  \emph{General Relativity} 
  (University of Chicago, 1984).

\bibitem{Medved}
  A. J. M. {Medved}, D. {Martin}, M. {Visser}, 
  {Dirty black holes: Symmetries at stationary nonstatic horizons}, 
  Phys. Rev. D 
  {\bf 70}, 
  024009 (2004), 8 pages,
  arXiv:gr-qc/0403026.

\bibitem{Zasl16}
  O. B. Zaslavskii,
  {Is the super-Penrose process possible near black holes?}
  {Phys. Rev. D} {\bf 93},
  {024056}
  (2016), 7 pages,
  arXiv:1511.07501 [gr-qc].  

\bibitem{Zasl11c}
  O. B. {Zaslavskii}, 
  {Acceleration of particles by black holes: Kinematic explanation}.
  Phys. Rev. D {\bf 84},
  024007 (2011), 6 pages,
  arXiv:1104.4802 [gr-qc].

\bibitem{Jacobson11}
  T. {Jacobson}, 
  {Where is the extremal Kerr ISCO?}
  Classical Quant. Grav. {\bf 28}, 
  187001 (2011), 5 pages,
  arXiv:1107.5081 [gr-qc].
   
\bibitem{HK11a}
  T. Harada, M. Kimura,
  {Collision of an innermost stable circular orbit particle around a Kerr black hole},
  {Phys. Rev. D} {\bf 83},
  {024002}
  (2011), 11 pages,
  arXiv:1010.0962 [gr-qc].
  
\bibitem{Wald74}
  R. M. {Wald}, 
  {Black hole in a uniform magnetic field}, 
  {Phys. Rev. D} 
  {\bf 10}, 1680-1685 
  (1974).

\bibitem{Hiscock81}
  W. A. {Hiscock}, 
  {On black holes in magnetic universes}, 
  J. Math. Phys. {\bf 22}, 1828-1833 
  (1981).

\end{thebibliography}
